\documentclass{aa}

\usepackage{natbib}
\usepackage{graphicx}
\usepackage{amsmath}
\usepackage{amsbsy}
\usepackage{txfonts}
\usepackage{fixltx2e}

\usepackage{xspace}
\xspaceaddexceptions{\@}

\usepackage{color}
\definecolor{Red}{rgb}{1.0,0.0,0.0}

\newcommand{\ten}[1]{$10^{#1}$}
\newcommand{\scit}[2]{$#1\times10^{#2}$}
\newcommand{\scim}[2]{#1\times10^{#2}}
\newcommand{\per}[1]{#1$^{-1}$}
\newcommand{\ps}{s$^{-1}$\xspace}
\newcommand{\pcs}{cm$^{-2}$\xspace}
\newcommand{\pcc}{cm$^{-3}$\xspace}

\newcommand{\micron}{$\mu$m\xspace}
\newcommand{\eq}[1]{Eq.\ \ref{eq:#1}}

\newcommand{\rx}[1]{Reaction \ref{rx:#1}}
\newcommand{\fig}[1]{Fig.\ \ref{fig:#1}}
\newcommand{\figg}[1]{Figure \ref{fig:#1}}
\newcommand{\tb}[1]{Table \ref{tb:#1}}
\newcommand{\sect}[1]{Sect.\ \ref{sec:#1}}
\newcommand{\sectt}[1]{Section \ref{sec:#1}}
\newcommand{\apx}[1]{Appendix \ref{apx:#1}}

\newcommand{\mh}{H$_2$\xspace}
\newcommand{\mhstar}{H$_2^\ast$\xspace}
\newcommand{\w}{H$_2$O\xspace}
\newcommand{\ctw}{$^{12}$C\xspace}
\newcommand{\cth}{$^{13}$C\xspace}
\newcommand{\nq}{$^{14}$N\xspace}
\newcommand{\nc}{$^{15}$N\xspace}
\newcommand{\twco}{$^{12}$CO\xspace}
\newcommand{\thco}{$^{13}$CO\xspace}
\newcommand{\nnhp}{N$_2$H$^+$\xspace}
\newcommand{\amh}{NH$_3$\xspace}
\newcommand{\mn}{N$_2$\xspace}
\newcommand{\mnq}{$^{14}$N$_2$\xspace}
\newcommand{\mnc}{N$^{15}$N\xspace}
\newcommand{\nqcr}{$^{14}$N/$^{15}$N ratio\xspace}
\newcommand{\nqcrs}{$^{14}$N/$^{15}$N ratios\xspace}

\newcommand{\cnq}{C$^{14}$N\xspace}
\newcommand{\cnc}{C$^{15}$N\xspace}
\newcommand{\hcnq}{HC$^{14}$N\xspace}
\newcommand{\hcnc}{HC$^{15}$N\xspace}
\newcommand{\htcn}{H$^{13}$CN\xspace}

\newcommand{\rxplus}{\ +\ }
\newcommand{\rxto}{\ \to\ }
\newcommand{\rxtoa}{\ \to\ &\ }
\newcommand{\ngas}{n_{\rm gas}\xspace}

\newcommand{\eup}{E_{\rm u}\xspace}

\begin{document}

\title{Nitrogen isotope fractionation in protoplanetary disks}

\author{
Ruud Visser\inst{1}
\and Simon Bruderer\inst{2}
\and Paolo Cazzoletti\inst{2}
\and Stefano Facchini\inst{2}
\and Alan N. Heays\inst{3}
\and Ewine F. van Dishoeck\inst{4,2}
}

\institute{
European Southern Observatory, Karl-Schwarzschild-Stra{\ss}e 2, 85748, Garching, Germany\\
\email{ruudvisser@gmail.com}
\and
Max-Planck-Institut f\"ur extraterrestrische Physik, Giessenbachstra{\ss}e 1, 85748 Garching, Germany
\and
Observatoire de Paris, LERMA, UMR 8112 du CNRS, 92195 Meudon, France
\and
Leiden Observatory, Leiden University, P.O.\ Box 9513, 2300 RA Leiden, The Netherlands
}

\date{Accepted on February 8, 2018}

\abstract
{} 
{The two stable isotopes of nitrogen, \nq and \nc, exhibit a range of abundance ratios both inside and outside the solar system. The elemental ratio in the solar neighborhood is 440. Recent ALMA observations showed HCN/\hcnc ratios from 83 to 156 in six T Tauri and Herbig disks and a CN/\cnc ratio of $323\pm30$ in one T Tauri star. We aim to determine the dominant mechanism responsible for these enhancements of \nc: low-temperature exchange reactions or isotope-selective photodissociation of \mn.} 
{Using the thermochemical code DALI, we model the nitrogen isotope chemistry in circumstellar disks with a 2D axisymmetric geometry. Our chemical network is the first to include both fractionation mechanisms for nitrogen. The model produces abundance profiles and isotope ratios for several key N-bearing species. We study how these isotope ratios depend on various disk parameters.} 
{The formation of CN and HCN is closely coupled to the vibrational excitation of \mh in the UV-irradiated surface layers of the disk. Isotope fractionation is completely dominated by isotope-selective photodissociation of \mn. The column density ratio of HCN over \hcnc in the disk's inner 100 au does not depend strongly on the disk mass, the flaring angle or the stellar spectrum, but it is sensitive to the grain size distribution. For larger grains, self-shielding of \mn becomes more important relative to dust extinction, leading to stronger isotope fractionation. Between disk radii of $\sim$50 and 200 au, the models predict HCN/\hcnc and CN/\cnc abundance ratios consistent with observations of disks and comets. The HCN/\hcnc and CN/\cnc column density ratios in the models are a factor of 2--3 higher than those inferred from the ALMA observations.} 
{} 


\maketitle


\section{Introduction}
\label{sec:intro}
Nitrogen has two stable isotopes, \nq and \nc, whose abundance ratio spans a wide range of values in the solar system: 50--280 in meteorites, 120--300 in interplanetary dust particles, $\sim$150 in comets, 272 on Earth and 440 in the solar wind \citep{busemann06a,floss06a,manfroid09a,marty11a,mumma11a,furi15a}. The solar wind value is representative of the gas out of which the solar system formed, so there must have been one or more mechanisms that acted to enhance \nc relative to \nq on Earth and in the other solid bodies.

Nitrogen isotope fractionation is also observed outside the solar system, from diffuse clouds to prestellar cores and circumstellar disks \citep{lucas98a,wampfler14a,zeng17a,guzman17a,hilyblant17a}. The lowest and highest \nqcr are a factor of 40 apart, clearly indicating various degrees of fractionation towards different sources. These observations were made in HCN, HNC, CN, \amh and \nnhp, which are all trace species of nitrogen. The bulk reservoir likely consists of atomic N and \mn, of which the isotopic composition remains unknown.

Two fractionation mechanisms have been proposed to explain the \nq/\nc ratios inside and outside the solar system: low-temperature isotope exchange reactions \citep{terzieva00a,roueff15a,wirstrom18a} and isotope-selective photodissociation of \mn \citep{liang07a,heays14a}. The low-temperature pathway is based on the difference in zero-point vibrational energies, such that heavier isotopologs are energetically favored over lighter isotopologs. This mechanism is well known for hydrogen and deuterium, leading e.g.\ to HDO/\w and DCO$^+$/HCO$^+$ abundance ratios of up to a few orders of magnitude higher than the elemental D/H ratio \citep{ceccarelli14a}. The isotope-selective dissociation pathway, also called self-shielding, arises from a small shift in frequencies for the absorption lines through which $^{14}$N$_2$ and $^{14}$N$^{15}$N are photodissociated. This shift allows photodissociation of $^{14}$N$^{15}$N to continue when the lines of $^{14}$N$_2$ are saturated, causing an enhancement of atomic \nc over \nq. The same mechanism, acting in CO, has been suggested as an explanation for the oxygen isotope ratios in meteorites \citep{lyons05a}.

Chemical models of nitrogen isotope fractionation have been published for dense clouds and prestellar cores \citep{terzieva00a,rodgers04a,rodgers08b,rodgers08a,wirstrom12a,roueff15a,wirstrom18a}, in all cases featuring low-temperature exchange reactions as the only fractionation mechanism. These models are generally successful at reproducing the fractionation observed in \amh and HCN, but have trouble reproducing some of the data on \nnhp. Using estimated \mn self-shielding functions and an incomplete chemical network, \citet{lyons09c} obtained an \hcnq/\hcnc ratio of $\sim$350 in a simplified circumstellar disk model. With new shielding functions and a larger network, \citet{heays14a} saw \hcnq/\hcnc ratios down to 50 in models of a single vertical cut through a disk. The observed range of \hcnq/\hcnc column density ratios is $83\pm32$ to $156\pm71$ in a sample of six T Tauri and Herbig disks \citep{guzman17a}. For \cnq/\cnc, there is a single measurement of $323\pm20$ in a T Tauri disk \citep{hilyblant17a}.

We present here the first nitrogen isotope fractionation models for circumstellar disks with a 2D axisymmetric geometry, including both low-temperature exchange reactions and self-shielding of \mn. We employ the thermochemical code DALI \citep[Dust and Lines;][]{bruderer12a,bruderer13a} in a setup similar to that used by \citet{miotello14a} to study the isotopologs of CO\@. \sectt{model} contains a full description of the physical framework and the chemical network. Abundance profiles and isotope ratios for a fiducial disk model are presented in \sectt{res}. We pay particular attention to the three cyanides HCN, HNC and CN because of the aforementioned recent observations of \nqcrs in HCN and CN \citep{guzman17a,hilyblant17a}. \sectt{disc} discusses the dominant fractionation mechanism, the match with observations, and the effects of changing various disk parameters. We summarize the main conclusions in \sectt{conc}.


\section{Model}
\label{sec:model}
We simulate the nitrogen isotope chemistry in a set of protoplanetary disk models with the thermochemical code DALI \citep[Dust and Lines;][]{bruderer12a,bruderer13a}. Given a two-dimensional, axisymmetric density profile, DALI uses a Monte Carlo technique to compute the radiation field and dust temperature throughout the disk. The gas temperature is solved by iterating over the heating/cooling balance and the chemical abundances. DALI approximates the excitation of the main coolants through an escape probability approach, taking into account both collisions and radiative effects. Once the gas temperature and abundance calculations are completed, a raytracer synthesizes continuum and line observations at arbitrary distances and inclinations. The accuracy of DALI has been verified against benchmark problems and observations \citep{bruderer12a,bruderer14a,bruderer13a,fedele13a}.

Our methods are generally the same as those developed by \citet{miotello14a} for CO and its isotopologs, except the focus has shifted to N-bearing species. The following subsections summarize the physical framework and describe the revised chemical network.


\subsection{Physical framework}
\label{sec:physmod}
The purpose of this work is to present a general exploration of the nitrogen isotope chemistry in protoplanetary disks, without focusing on any one particular source. Isotope effects depend primarily on the gas temperature and the UV flux, so we adopt a small grid of parameterized models to cover a range of values for both quantities (Tables \ref{tb:modpar} and \ref{tb:stespec}).

\begin{table}[t!]
\caption{Summary of model parameters.}
\label{tb:modpar}
\centering
\begin{tabular}{cc}
\hline\hline
Parameter & Value(s) \\
\hline
$R_{\rm in}$ & 0.07 au \\
$R_{\rm c}$ & 60 au \\
$R_{\rm out}$ & 600 au \\
$\gamma$ & 1 \\
$h_{\rm c}$ & 0.1 rad \\
$\psi$ & 0.1, \textit{0.2}, 0.3 \\
$\chi$ & 0.2 \\
$f_{\rm L}$ & 0.1, \textit{0.9}, 0.99 \\
$M_{\rm gas}$ & \ten{-4}, $\mathit{10^{-3}}$, \ten{-2} $M_\odot$ \\
$M_{\rm gas}$/$M_{\rm dust}$ & 100 \\
$\begin{array}{c}{\rm stellar\ spectrum}\\({\rm see\ Table\ \ref{tb:stespec}\ and\ text})\\\end{array}$ & $\left\{ \begin{array}{l}4\,000\ {\rm K},\ 1\ L_\odot,\ \dot{M}=10^{-9}\ M_\odot\ \rm{yr}^{-1};\\\mathit{4\,000\ K,\ 1\ L_\odot,\ \dot{M}=10^{-8}\ M_\odot\ \mathit{yr}^{-1}};\\10\,000\ {\rm K},\ 10\ L_\odot,\ \dot{M}=0\end{array}\right.$ \\
$L_{\rm X}$ & \ten{30} erg \ps \\
$\zeta_{{\rm H}_2}$ & \scit{5}{-17} \ps \\
\hline
\end{tabular}
\tablefoot{Values in italics are used in the fiducial model.}
\end{table}

\begin{table}[t!]
\caption{Luminosities integrated over various wavelength ranges for the model spectra.}
\label{tb:stespec}
\centering
\begin{tabular}{cccccc}
\hline\hline
Type & $\dot{M}$            & $L_\ast$    & $L_{\rm acc}$ & $L_{\rm 6-13.6\ eV}$ & $L_{\rm 912-1000\ \textup{\AA}}$ \\
     & ($M_\odot$ \per{yr}) & ($L_\odot$) & ($L_\odot$) & ($L_\odot$) & ($L_\odot$) \\
\hline
Herbig  & 0        & 10  & 0         & $7.7(-1)$ & $2.2(-3)$\rule{0pt}{0.9em} \\
T Tauri & $1(-8)$  & 1.0 & $2.3(-1)$ & $1.8(-2)$ & $5.0(-5)$ \\
T Tauri & $1(-9)$  & 1.0 & $2.3(-2)$ & $1.8(-3)$ & $5.0(-6)$ \\
T Tauri & $1(-10)$ & 1.0 & $2.3(-3)$ & $2.0(-4)$ & $5.0(-7)$ \\
T Tauri & 0        & 1.0 & 0         & $2.7(-5)$ & $1.8(-12)$ \\
\hline
\end{tabular}
\tablefoot{The notation $a(-b)$ means \scit{a}{-b}.}
\end{table}

The disk models have a power-law surface density ($\Sigma \propto R^{-\gamma}$) with an exponential taper beyond $R_{\rm c}=60$ au \citep{andrews11a}. The gas densities follow a vertical Gaussian profile with a scale height angle $h = h_{\rm c} (R/R_{\rm c})^\psi$, where $\psi$ varies from 0.1 to 0.3 to cover a range of flaring angles. The dust consists of populations of small grains (0.005--1 \micron) and large grains \citep[1-1000 \micron;][]{dalessio06a}. The small grains have the same scale height $h$ as the gas, while the large grains have settled to a reduced scale height of $\chi h$ with $\chi=0.2$. We explore different grain size distributions by varying the mass fraction of large grains ($f_{\rm L}$) from 10\% to 99\%. The global gas-to-dust mass ratio is constant at 100. Within the grid of models, the total disk mass varies between \ten{-4}, \ten{-3} and \ten{-2} $M_\odot$. All other parameters are set to the same values as in \citet{miotello14a}, including a cosmic ray ionization rate of \scit{5}{-17} \ps per \mh, a stellar X-ray luminosity of \ten{30} erg \ps, and an interstellar UV flux typical for the solar neighborhood \citep{habing68a}.

The stellar spectra considered in our models consist of a blackbody spectrum of a given temperature and luminosity: 4000 K and 1 $L_\odot$ for a T Tauri star and 10\,000 K and 10 $L_\odot$ for a Herbig Ae/Be star \citep[the same as in][]{miotello14a}. The T Tauri spectra include an optional UV excess due to accretion. For a given accretion rate $\dot{M}$, we convert all gravitational potential energy into 10\,000 K blackbody radiation emitted uniformly over the stellar surface. The accretion luminosity is
\begin{equation}
L_{\rm acc} = \frac{G M_\ast \dot{M}}{R_\ast}\,.
\end{equation}
The median accretion rate for T Tauri stars of 1 Myr old is \ten{-8} $M_\odot$ \per{yr} \citep{hartmann98a}. Setting $M_\ast = 1$ $M_\odot$ and $R_\ast = 1.5$ $R_\odot$, the accretion luminosity is 0.23 $L_\odot$ integrated across all possible wavelengths or 0.018 $L_\odot$ in the far-UV range from 6 to 13.6 eV.

\begin{figure}[t!]
\centering
\includegraphics[width=\hsize]{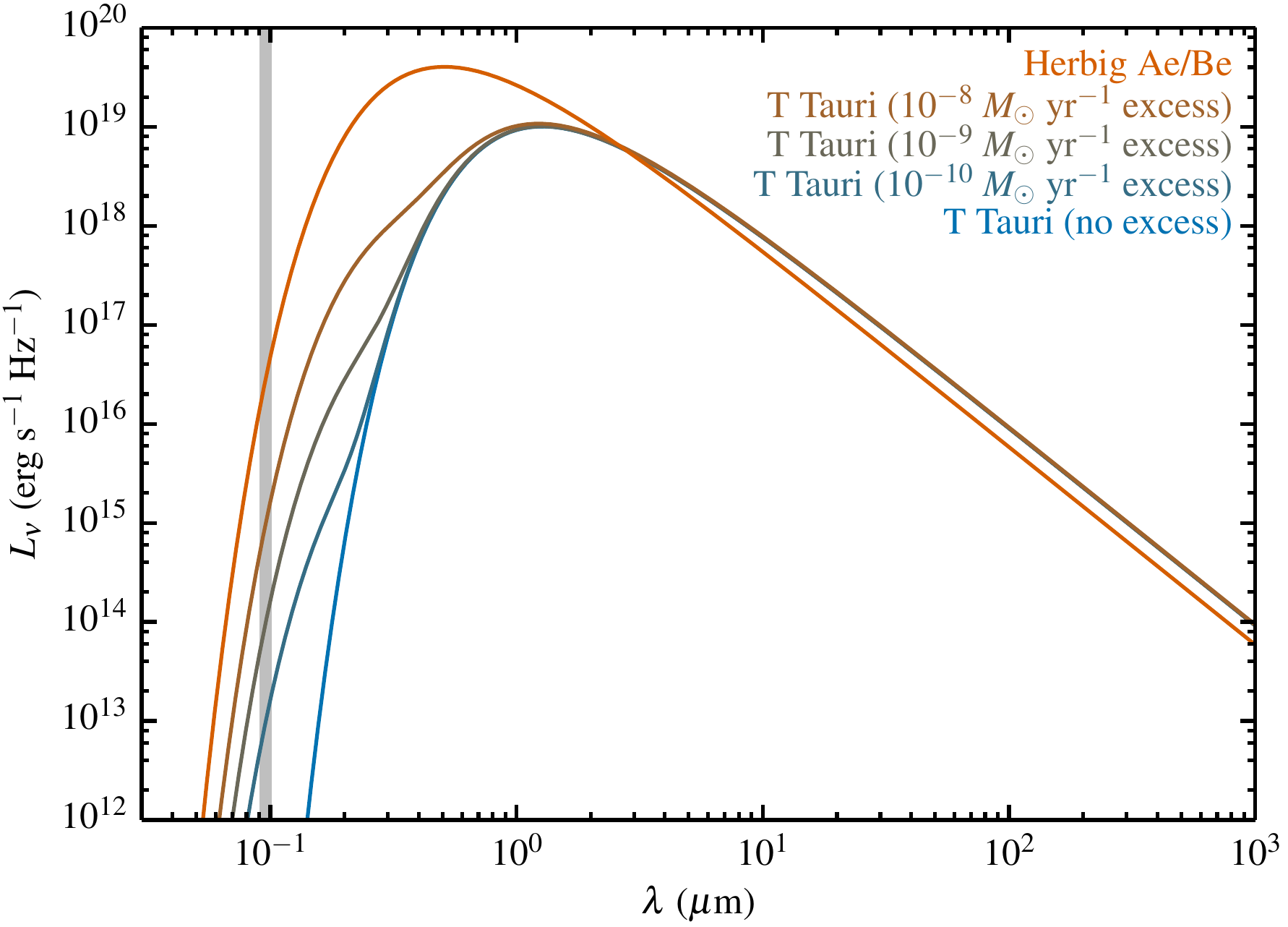}
\caption{Stellar spectra used in our model. The gray band marks the wavelength range of 912 to 1000 {\AA} for the photodissociation of \mn.}
\label{fig:stespec}
\end{figure}

\figg{stespec} shows the stellar spectra for a T Tauri star without UV excess, three T Tauri stars with different accretion rates, and a Herbig star. These spectra vary by orders of magnitude in flux between 912 and 1000 \AA, the wavelength range critical for the photodissociation and self-shielding of \mn \citep{li13a,heays14a}. \tb{stespec} lists the luminosities integrated over various wavelength ranges.

\subsection{Chemical network}
\label{sec:chemnet}


\subsubsection{Reduced network for C/N/O chemistry}
\label{sec:rednet}
For their exploration of the CO isotopolog chemistry, \citet{miotello14a} extended the reaction network of \citet{bruderer12a} to include \cth, $^{17}$O and $^{18}$O\@. The \citeauthor{bruderer12a} network itself is a reduced version of the UMIST Database for Astrochemistry \citep[UDfA;][]{woodall07a,mcelroy13a}, optimized to yield accurate abundances for CO and related species with a minimum of computational effort. However, it lacks several species and reactions crucial to the chemistry of HCN, HNC and CN \citep[see review by][]{loison14a}.

Our model combines the reaction networks of \citet{bruderer12a} and \citet{loison14a} into a new reduced network for the C/N/O chemistry throughout a protoplanetary disk. Cross-checks against the full UMIST database at a dozen sets of physical conditions (representative of different physical regimes within the disk) confirmed that our network contains all reactions that affect the abundances of HCN, HNC and CN by more than a few per cent. Any effects smaller than this margin are lost in the overall model uncertainties.

The network contains both gas-phase and grain-surface processes, as described in Appendix A.3.1 of \citet{bruderer12a}. The gas-phase part consists of standard neutral-neutral and ion-molecule chemistry; reactions with vibrationally excited \mh; charge exchange and H transfer with polycyclic aromatic hydrocarbons; photodissociation and photoionization; and reactions induced by X-rays and cosmic rays. All neutral molecules can freeze out, and desorption back into the gas phase can occur thermally or non-thermally via UV photons and cosmic rays. Grain-surface chemistry is limited to \mh formation, hydrogenation of CN to HCN, and step-wise hydrogenation of atomic C, N and O to CH$_4$, \amh and H$_2$O \citep{bruderer12a}. We adopt desorption energies from the Kinetic Database for Astrochemistry \citep[KIDA;][]{wakelam12a}, including 1600 K for CN, 2050 K for HCN and HNC, and 5500 K for \amh.

The interaction of cosmic rays or X-rays with \mh generates a local UV field even in parts of the disk that are shielded from stellar and interstellar UV radiation. The UDfA rate equation for the CR-- and X-ray--induced photodissociation of CO is a fit to a set of tabulated values from \citet{gredel87a}. These authors reported faster rates for higher temperatures, but the new calculations from \citet{heays14a} show no dependence on temperature. The \citeyear{gredel87a} calculations appear not to have accounted for \mh shielding, or at least not for the effect of the gas temperature on the \mh shielding.

The temperature-independent rate coefficient $k$ for this process can be expressed as the product of the total ionization rate $\zeta$ (sum of cosmic ray and X-ray contributions) and the photodissociation efficiency $P$\@. To account for self-shielding, we fit a sigmoid function to the tabulated values of \citeauthor{heays14a} and some additional values for typical interstellar grain properties:
\begin{equation}
P = \frac{56.14}{1 + 51100 \left[x({\rm CO})\right]^{0.792}} + 4.3\,,
\end{equation}
where $x({\rm CO}) \equiv n({\rm CO})/\ngas$ is the CO abundance relative to the total number density of gas. The photodissociation efficiency varies smoothly from 5 at a CO abundance of a few \ten{-4} to 60 at abundances below \ten{-8}. Grain growth can increase $P$ by up to a factor of 2 \citep{heays14a}.

Our network includes vibrationally excited \mh in a two-level approximation, where \mhstar denotes a pseudo-level with an energy of 30\,163 K \citep{london78a}. This energy can be used to overcome activation barriers as detailed in Appendix A.3.1 of \citet{bruderer12a}. An important example for the cyanide chemistry is the reaction between N and \mh, whose barrier of 12\,650 K \citep{davidson90a} would otherwise be insurmountable in the bulk of the disk. In our model, FUV pumping of \mh to \mhstar in the disk's surface layers enables the reaction with N, and the product NH enters directly into the cyanide chemistry by reacting with C or C$^+$ to form CN or CN$^+$ (\apx{formdest}). \apx{h2star} describes the treatment of \mhstar in more detail.

We run the chemistry for a typical disk lifetime of 1 Myr. The initial abundances are listed in \tb{initabun}. These values are adopted from \citet{cleeves15a}, except $x({\rm CO})$ is set to \scit{1}{-4} relative to total hydrogen.

\begin{table}[t!]
\caption{Initial abundances relative to the total number of hydrogen atoms.}
\label{tb:initabun}
\centering
\begin{tabular}{cccc}
\hline\hline
Species & Abundance & Species & Abundance \\
\hline
H$_2$ & $5.000(-01)$ & $^{13}$CN & $9.565(-10)$ \\
He & $1.400(-01)$ & $^{13}$C$^{15}$N & $2.126(-12)$ \\
H$_2$O ice & $2.500(-04)$ & HCN & $1.000(-08)$ \\
CO & $1.000(-04)$ & HC$^{15}$N & $2.222(-11)$ \\
$^{13}$CO & $1.449(-06)$ & H$^{13}$CN & $1.449(-10)$ \\
NH$_3$ ice & $9.900(-06)$ & H$^{13}$C$^{15}$N & $3.220(-13)$ \\
$^{15}$NH$_3$ ice & $2.200(-08)$ & H$_3^+$ & $1.000(-08)$ \\
N & $5.100(-06)$ & HCO$^+$ & $9.000(-09)$ \\
$^{15}$N & $1.133(-08)$ & H$^{13}$CO$^+$ & $1.304(-10)$ \\
N$_2$ & $1.000(-06)$ & C$_2$H & $8.000(-09)$ \\
N$^{15}$N & $4.444(-09)$ & $^{13}$CCH & $1.159(-10)$ \\
C & $7.000(-07)$ & C$^{13}$CH & $1.159(-10)$ \\
$^{13}$C & $1.014(-08)$ & C$^+$ & $1.000(-09)$ \\
PAH & $6.000(-07)$ & $^{13}$C+ & $1.449(-11)$ \\
CH$_4$ & $1.000(-07)$ & Mg$^+$ & $1.000(-11)$ \\
$^{13}$CH$_4$ & $1.449(-09)$ & Si$^+$ & $1.000(-11)$ \\
CN & $6.600(-08)$ & S$^+$ & $1.000(-11)$ \\
C$^{15}$N & $1.467(-10)$ & Fe$^+$ & $1.000(-11)$ \\
\hline
\end{tabular}
\end{table}


\subsubsection{Addition of $^\mathsf{15}$N and $^\mathsf{13}$C}
\label{sec:addiso}
Observationally, the \nqcr is often obtained from the optically thin emission of \htcn and \hcnc\@. Analyzing the nitrogen isotope chemistry therefore requires the addition of both \nc and \cth to the network. The oxygen isotopes are not necessary. We extended the reduced C/N/O network according to standard procedure \citep[e.g.,][]{lebourlot93a,rollig13a}. This involves single substitutions of either \nq by \nc or \ctw by \cth, or substitution of both \nq and \ctw, but not double substitutions of the same atom. For example, \nnhp expands into $^{14}$N$_2$H$^+$, $^{14}$N$^{15}$NH$^+$ and $^{15}$N$^{14}$NH$^+$. Our species of primary interest, HCN, becomes H$^{12}$C$^{14}$N, H$^{12}$C$^{15}$N, H$^{13}$C$^{14}$N and H$^{13}$C$^{15}$N\@. \tb{speclist} in \apx{speclist} lists the full set of species in the network.

The reactions from the C/N/O network are expanded to include all possible isotope substitutions, under the condition that functional groups are preserved \citep{woods09a}. For example, the proton-transfer reaction \hcnc + \nnhp can only produce HC$^{15}$NH$^+$ + \mn, not HCNH$^+$ + N$^{15}$N\@. If the reaction mechanism is unknown, then we follow the guidelines from \citet{rollig13a}: minimize the number of bonds broken and formed; favor transfer of H or H$^+$ over heavier atoms; and favor destruction of the weaker of two possible bonds. In the absence of experimental data, the rate constants for all isotopolog reactions are set to be the same as for the original reaction. In cases where multiple product channels are allowed, all channels are assigned an equal probability \citep{rollig13a}.


\subsubsection{Isotope-exchange reactions}
\label{sec:iso-ex}
At temperatures below $\sim$20 K, isotope-exchange reactions tend to enhance the abundances of heavier isotopologs over their lighter counterparts. The most common astrophysical example is the exothermic reaction between $^{13}$C$^+$ and CO to form C$^+$ and \thco \citep{watson76a}. Its forward rate at low temperatures is faster than that of the return reaction, leading to carbon isotope fractionation in CO\@. Similar reactions can affect the \nqcr in HCN and other species.

\citet{roueff15a} reviewed a set of 19 possible carbon and nitrogen isotope-exchange reactions, of which eleven were deemed to be viable in astrophysical environments: numbers 1, 2, 3, 4, 5, 12, 14, 15, 16, 17 and 19 in their Table 1. Our chemical network includes these eleven reactions, with the rates and barriers provided by \citeauthor{roueff15a} Although these rates appear to be the best currently available, they may have been underestimated for some reactions \citep{wirstrom18a}.


\subsubsection{Isotope-selective photodissociation}
\label{sec:isophotdis}
Another mechanism to alter isotopolog ratios is isotope-selective photodissocation. CO and \mn can dissociate only through absorption into a discrete set of narrow far-UV lines \citep{visser09b,li13a}. These lines become saturated for CO and \mn column densities of $\gtrsim$\ten{14} \pcs, at which point self-shielding sets in: molecules deep inside a disk or cloud are shielded from dissociating radiation by molecules near the surface.

The rarer isotopologs \thco and \mnc absorb photons at different frequencies than the primary species, and these photons are less strongly shielded. At any depth into a disk or cloud, \thco and \mnc therefore dissociate faster than \twco and \mnq. Through additional reactions, this affects the \nqcr in the cyanides (\sect{14n15n}). Our network includes the CO and \mn shielding functions from \citet{visser09b} and \citet{heays14a}.\footnote{http://home.strw.leidenuniv.nl/$\sim$ewine/photo} From the available excitation temperatures and Doppler widths, the best choices are 20 K and 0.30 km \ps for CO and 30 K and 0.13 km \ps for \mn. These data files include the effects of shielding by H and \mh, though our models show that mutual shielding is not important for nitrogen isotope fractionation in disks.

For a given point in the disk, DALI computes the shielding functions both radially towards the star and vertically towards the disk surface. The photodissociation rate is set by the direction that provides the smaller amount of shielding. The same 1+1D approximation was used by \citet{miotello14a}.


\section{Results}
\label{sec:res}
We present the key characteristics of the nitrogen isotope chemistry for a fiducial disk model of \ten{-3} $M_\odot$ with 90\% large grains and a scale height angle exponent of $\psi=0.2$. The disk is illuminated by a T Tauri stellar spectrum with a UV excess for an accretion rate of \ten{-8} $M_\odot$ \per{yr}, providing an FUV luminosity of 0.018 $L_\odot$ between 6 and 13.6 eV\@. The elemental isotope ratios are 440 for \nq/\nc and 69 for \ctw/\cth\@. \sectt{param} discusses how the chemistry depends on various model parameters.


\subsection{Physical characteristics}
\label{sec:physchar}
\begin{figure*}[t!]
\centering
\includegraphics[width=\hsize]{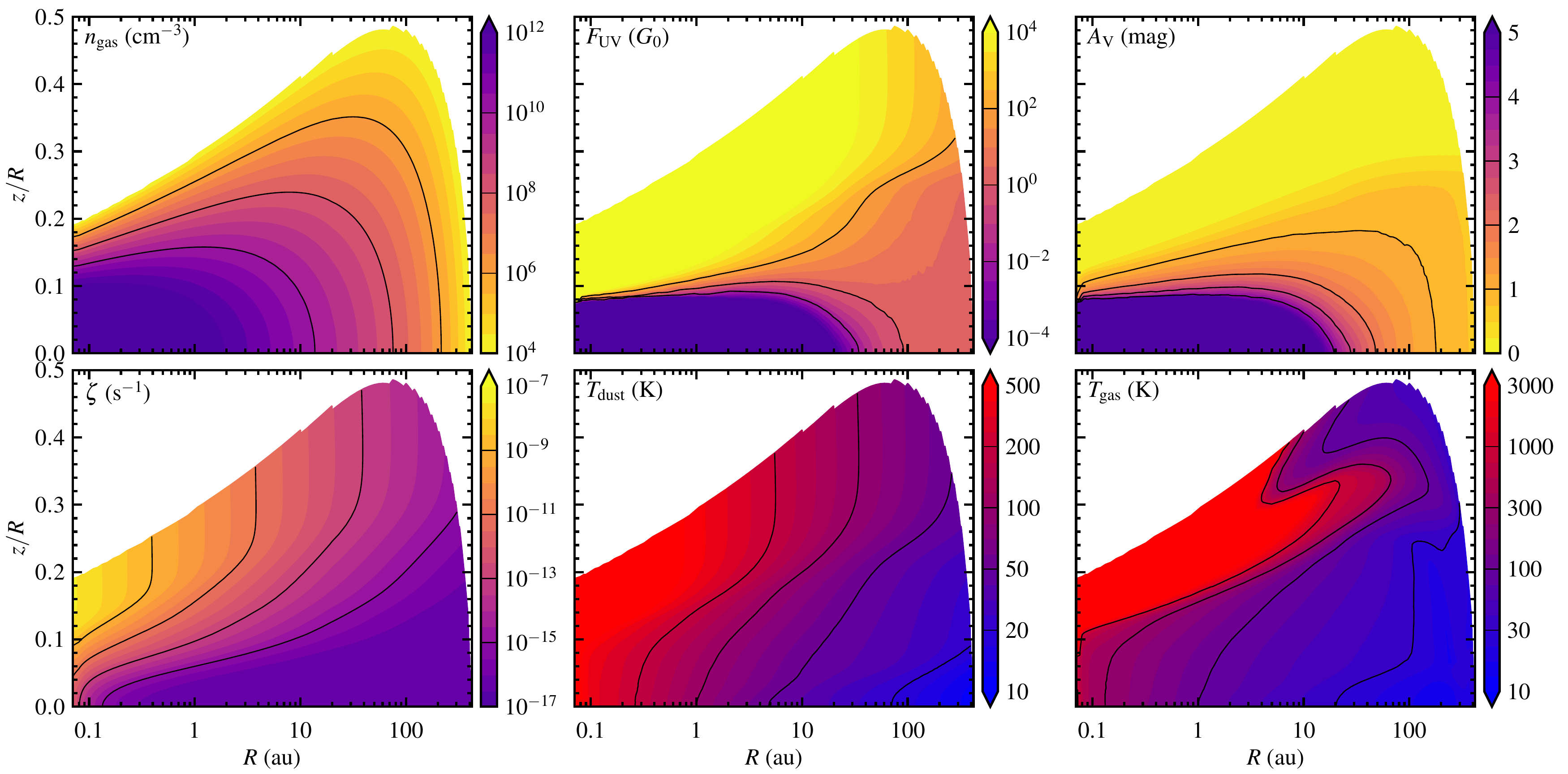}
\caption{Gas density, UV flux, visual extinction, ionization rate (sum of X-ray and cosmic ray contributions), dust temperature and gas temperature for a \ten{-3} $M_\odot$ disk with 90\% large grains, $\chi=0.2$ and $\psi=0.2$, illuminated by a T Tauri stellar spectrum with a UV excess for \ten{-8} $M_\odot$ \per{yr}. The $F_{\rm UV}$ plot accounts for extinction. All panels are truncated at the \ten{4} \pcc density contour.}
\label{fig:physprop}
\end{figure*}

\begin{figure*}[t!]
\centering
\includegraphics[width=\hsize]{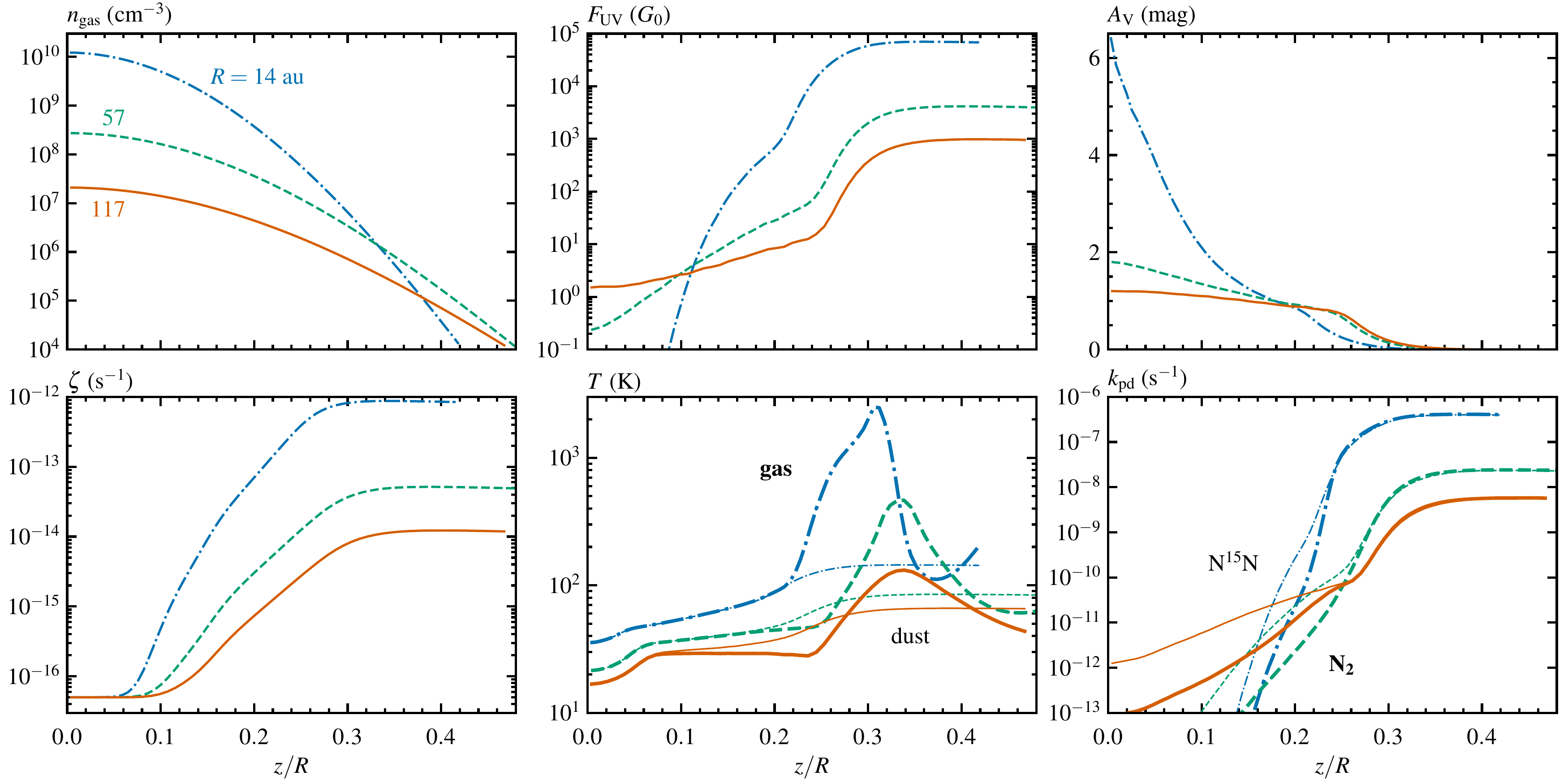}
\caption{Vertical cuts corresponding to \fig{physprop}. The bottom center panel shows the dust temperatures as thinner lines and gas temperatures as thicker lines. The bottom right panel shows the photodissociation rate of N$^{15}$N as thinner lines and that of \mn as thicker lines.}
\label{fig:vcutphys}
\end{figure*}

\figg{physprop} shows the two-dimensional profiles of the gas density, UV flux, visual extinction, ionization rate, dust temperature and gas temperature. Vertical cuts are plotted in \fig{vcutphys} for three radii, corresponding to the abundance maximum of HCN in the surface layers ($R=14$ au), the abundance maximum of CN (57 au) and the strongest N isotope fractionation in HCN (117 au).

The UV flux in both figures is plotted in units of the mean interstellar radiation field \citep[$G_0$;][]{habing68a} and includes effects of geometrical dilution and dust attenuation, as computed through the full 2D continuum radiative transfer. Due to scattered stellar light, the midplane UV flux reaches the equivalent of the interstellar field ($G_0=1$) at a radius of 95 au. The extinction profile is the result of comparing the actual intensity at each point in the disk to the intensity expected based only on geometrical dilution of the stellar radiation \citep{visser11a}.

Ionization is controlled by X-rays out to 0.3 au along the midplane, after which the X-ray ionization rate drops below the cosmic ray contribution of \scit{5}{-17} \ps. The midplane dust temperature decreases to 100 K at 1 au and 20 K at 70 au, marking the approximate locations of the \w and CO snowlines.

The bottom right panel of \fig{vcutphys} shows the photodissociation rates of N$^{15}$N (thinner lines) and \mn (thicker lines). Self-shielding of \mn is important below $z/R\approx0.25$ (where $A_{\rm V}\approx0.5$) and results in an order-of-magnitude reduction in the photodissociation rate relative to N$^{15}$N. The overall decrease of the photodissociation rates from the disk surface down to the midplane is due to extinction by dust. Shielding of \mn and N$^{15}$N by H or \mh is negligible.


\subsection{Cyanide abundances and morphologies}
\label{sec:abunprof}
\begin{figure*}[t!]
\centering
\includegraphics[width=\hsize]{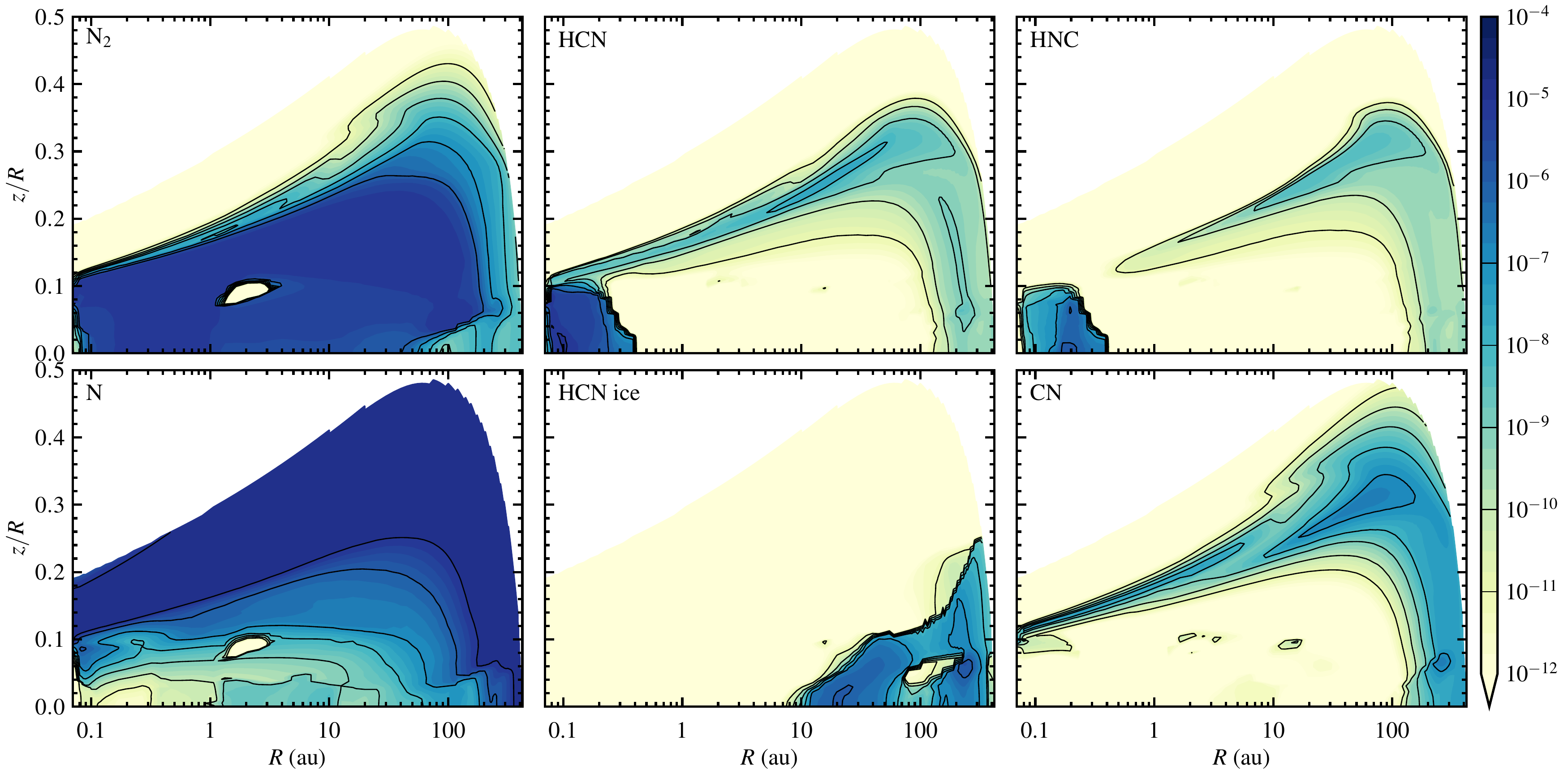}
\caption{Abundances at 1 Myr of several N-bearing molecules for the same model as in \fig{physprop}. Values are relative to $n_{\rm gas} \approx 2n({\rm H}_2) + n({\rm H})$. Contours are drawn at intervals of a factor of 10 from \ten{-11} up to \ten{-5}.}
\label{fig:abun6}
\end{figure*}

\begin{figure*}[t!]
\centering
\includegraphics[width=\hsize]{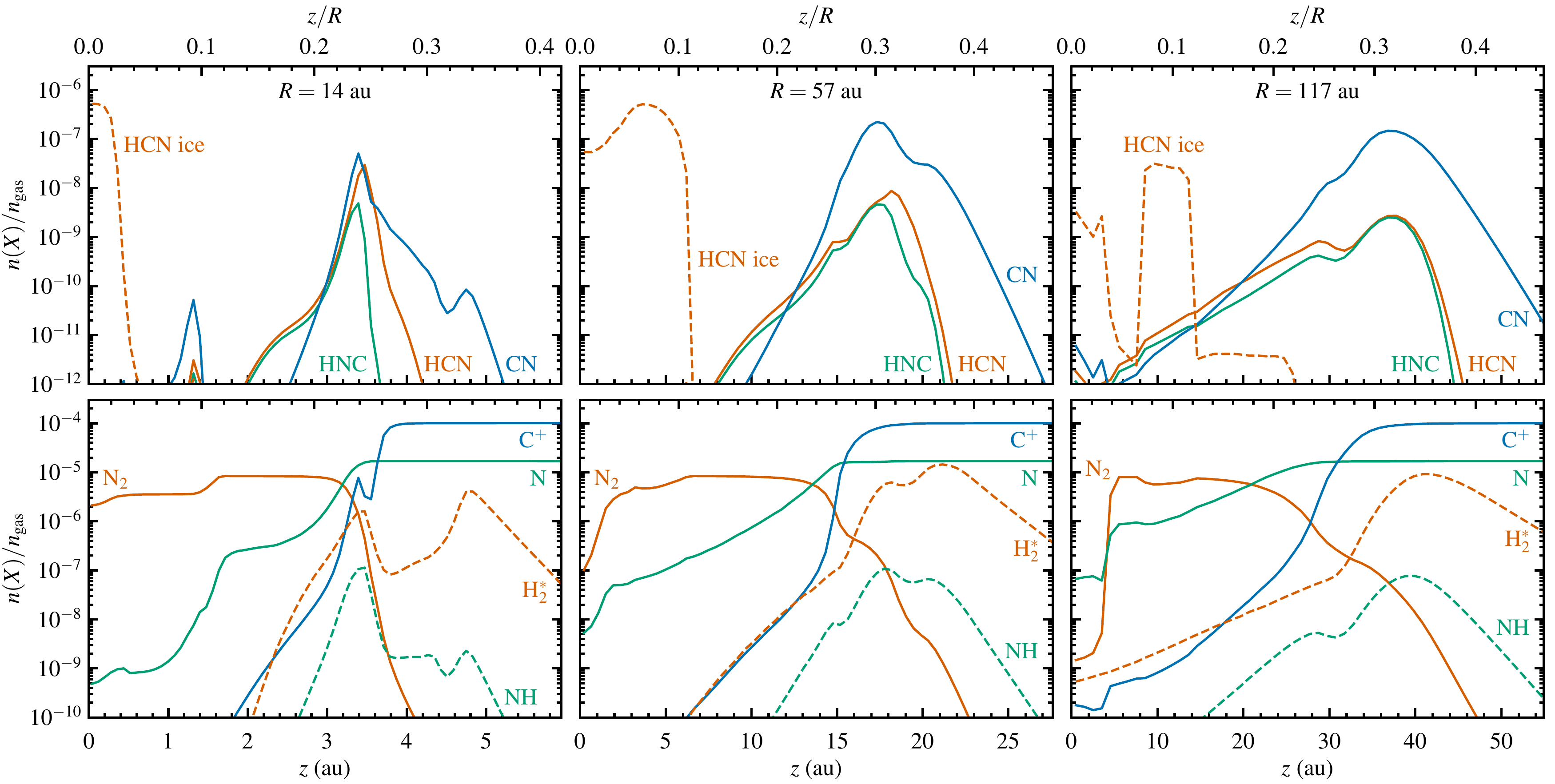}
\caption{Vertical cuts corresponding to \fig{abun6}.}
\label{fig:vcutabun}
\end{figure*}

The abundance profiles for \mn, N, HCN, HCN ice, HNC and CN at 1 Myr in the fiducial model are plotted in \fig{abun6} and the corresponding vertical cuts appear in \fig{vcutabun}. Across the entire disk, the dominant forms of nitrogen are \mn (47\%), \amh ice (30\%), N (8\%), \mn ice (7\%) and NO ice (6\%). HCN ice accounts for 1\% of all nitrogen and gas-phase HCN for only 0.08\%. At the inner edge, however, HCN is the second-most abundant N-bearing species (30\% for $R<0.2$ au).

The cyanide morphologies from \fig{abun6} are qualitatively similar to those seen in other models \citep[e.g.,][]{aikawa02a,markwick02a,jonkheid07a,agundez08a,woods09a,walsh10a,fogel11a,semenov11a,cleeves11a,podio14a}. The high HCN abundance in the inner disk is observed at near- and mid-infrared wavelengths \citep{lahuis06a,gibb07a,carr08a,salyk08a,salyk11a,mandell12a}, and the colder cyanides in the surface and outer disk are seen with millimeter and submillimeter telescopes \citep{dutrey97a,thi04a,salter11a,graninger15a,guzman15a,guzman17a,teague16a,hilyblant17a,vanterwisga17a}.

Aside from the high-abundance inner region, HCN shows a secondary maximum of \scit{3}{-8} in a thin layer along the disk surface. Near the midplane, HCN is absent from $R=0.4$ to 130 au. HCN ice is abundant (up to \scit{1}{-6}) outside 8 au. HNC follows the same pattern as HCN, but the HNC abundance is systematically lower by about an order of magnitude. For comparison, observations of one T Tauri disk and one Herbig disk showed HNC/HCN line intensity ratios of 0.1 to 0.2 \citep{graninger15a}. CN is confined to the surface layers and outer disk, where it is typically two orders of magnitude more abundant than HCN\@. CN is more spatially extended than HCN, consistent with ALMA observations \citep{guzman15a}. Furthermore, the CN morphology from \fig{abun6} gives rise to a ring of low-$J$ CN emission, as also observed with ALMA \citep{teague16a,vanterwisga17a}. These CN rings were analyzed in detail by \citet{cazzoletti17a}.

\apx{formdest} offers a complete breakdown of the various reactions that govern the formation and destruction of HCN, HNC and CN\@. The remainder of the current section describes the key physical and chemical aspects of the cyanide morphologies from Figs.\ \ref{fig:abun6} and \ref{fig:vcutabun}.

\textbf{Inner Disk} -- HCN is abundant in the inner 0.2 au, fueled by the combination of high temperatures (up to 1300 K) and high ionization rates (up to a few \ten{-11} \ps; \fig{physprop}). HCN is formed through all possible pathways starting from \amh, N and \mn (Reactions \ref{rx:amh/h3o+}--\ref{rx:n+/h2}). The conversion rate of HNC + H $\to$ HCN + H (\rx{hnc/h}) decreases with temperature, so the HNC abundance increases from the inner rim to 0.2 au. The CN abundance is low in this region, because CN + \mh $\to$ HCN + H (\rx{cn/h2}) drives all CN to HCN.

\textbf{Midplane} -- The midplane temperature drops below 230 K at 0.2 au, disabling the gas-phase formation of water through O + \mh $\to$ OH and OH + \mh $\to$ H$_2$O\@. Outside 0.2 au, the abundance of atomic O is high enough to drain the entire cyanide reservoir on a timescale of a few thousand years through CN + O $\to$ CO + N (\rx{cn/o}). The snowlines of HCN and HNC lie at 8 au and that of CN at 16 au. At larger radii, any cyanides formed in the gas freeze onto the dust and are safe from destruction by atomic O\@. Gas-phase cyanides remain absent along the midplane until 130 au, at which point UV-driven chemistry takes over.

\textbf{Surface} -- Photodissociation and photoionization rule the surface layers of the disk. The UV field dissociates \mn and CO, and it pumps \mh into a vibrationally excited state. This \mhstar reacts with N to NH, which reacts further with C or C$^+$ to produce HCN, HNC and CN (Reactions \ref{rx:n/h2*}--\ref{rx:nh/c+}). Most of the surface is warm enough to overcome the barrier of 960 K on CN + \mh $\to$ HCN + H (\rx{cn/h2}), so formation of HCN from CN and \mh is faster than photodissociation of HCN to CN and H\@. The two dominant cyanide loss channels are photodissociation of CN and the reaction of CN with O.

\textbf{Outer Disk} -- Photodissociation and photoionization remain important at intermediate altitudes and in the outer disk, but the UV flux is lower and the gas is colder than in the surface layers. The lower abundance of \mhstar slows down production of CN from N and NH (Reactions \ref{rx:n/h2*}--\ref{rx:nh/c+}). Production of HCN slows down even more because of the barrier on CN + \mh $\to$ HCN + H (\rx{cn/h2}).

To highlight the importance of vibrationally excited \mh for the cyanide abundances, consider the vertical cuts in \fig{vcutabun}. At each radius, CN, HCN, HNC all peak in a narrow band of altitudes near $z/R$ of 0.2 to 0.3. When the abundances are high enough, a shoulder or secondary peak appears at $z/R$ of 0.3 to 0.4. NH and \mhstar have the same dual-peaked profiles as the cyanides. In contrast, the abundances of N and C$^+$ follow an entirely different pattern.

\begin{figure*}[t!]
\centering
\includegraphics[width=\hsize]{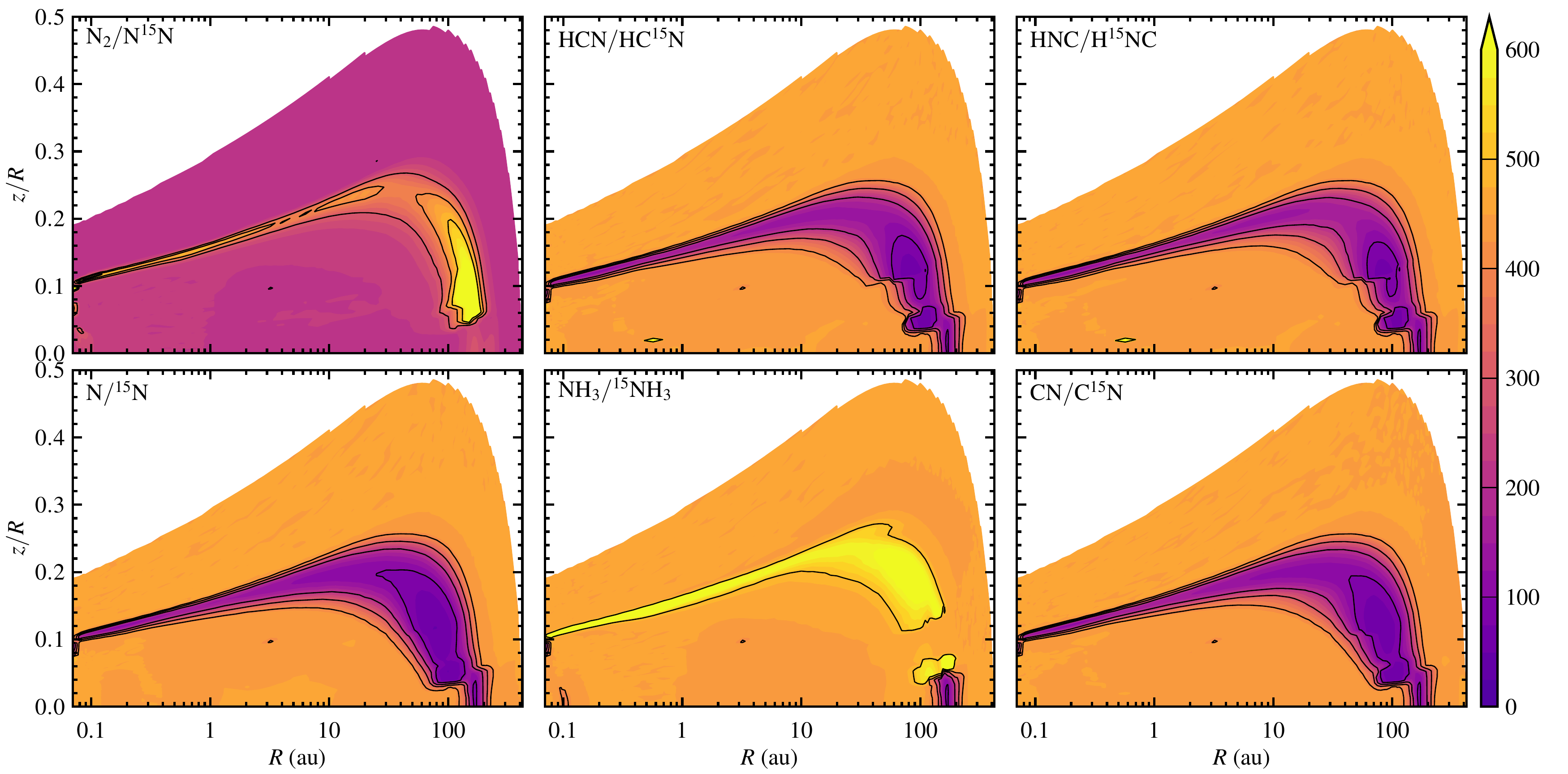}
\caption{Nitrogen isotope ratios of several molecules at 1 Myr for the same model as in \fig{physprop}. The elemental \nqcr is 440. Contours are drawn at ratios of 100, 200, 300, 400 and 500.}
\label{fig:ratio6n}
\end{figure*}

\begin{figure*}[t!]
\centering
\includegraphics[width=\hsize]{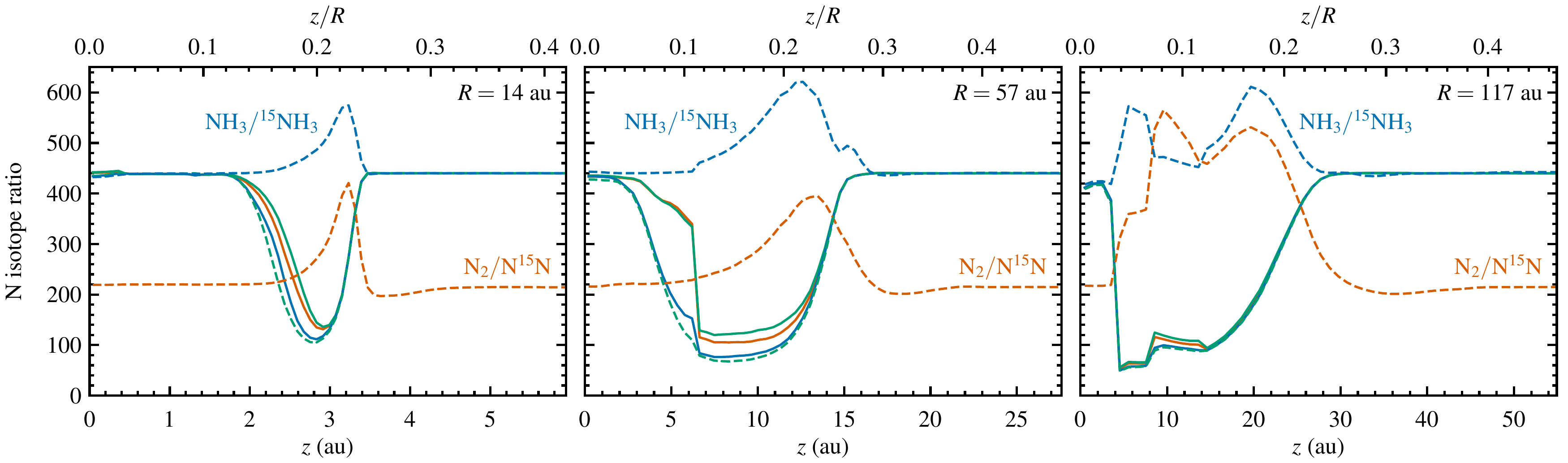}
\caption{Vertical cuts corresponding to \fig{ratio6n}. The profiles for \mn/N$^{15}$N and \amh/$^{15}$NH$_3$ are labeled. From top to bottom, the remaining profiles are for HNC/H$^{15}$NC (solid green), HCN/\hcnc (solid orange), CN/C$^{15}$N (solid blue) and N/\nc (dashed green).}
\label{fig:vcut-nratio}
\end{figure*}

Based on the similar vertical abundance profiles, the chemistry of \mhstar, NH and the cyanides is closely linked. This conclusion is also evident when plotting the abundances as a function of radius at constant $z$ or $z/R$ through the CN maximum (not shown). The abundance profile of \mhstar represents a balance between excitation due to UV pumping and deexcitation due to spontaneous decay and collisions with ground-state H or \mh. At altitudes above the CN maximum, photodissociation of \mh provides an additional loss channel for \mhstar.


\subsection{Isotope ratios}
\label{sec:iratios}


\subsubsection{$^\mathsf{14}$N/$^\mathsf{15}$N}
\label{sec:14n15n}
\figg{ratio6n} shows the nitrogen isotope ratios for \mn, N, HCN, HNC, CN and \amh in the fiducial disk model. Vertical cuts at 14, 57 and 117 au appear in \fig{vcut-nratio}. All six species undergo isotope fractionation in a region extending from near the disk surface down to the midplane at 100--200 au.

Molecular nitrogen is depleted in \nc, reaching a peak \mn/N$^{15}$N ratio of 750 at a radius of 150 au. This value is 3.4 times higher than the primordial \mn/N$^{15}$N ratio ratio of 220. Fractionation in \mn is fully driven by isotope-selective photodissociation, which lowers the photodissociation rate of \mn by an order of magnitude compared to N$^{15}$N (\fig{vcutphys}, bottom right). The relative importance of isotope-selective photodissociation and low-temperature isotope exchange reactions is discussed in more detail in \sect{mech}.

The isotope-selective photodissociation of \mn leads to strong fractionation in atomic N in the bottom left panel of \fig{ratio6n}. The N/$^{15}$N ratio reaches a minimum of 48 at $R=82$ au, 9.2 times lower than the initial ratio of 440. Formation of CN, HCN and HNC generally begins with atomic N (\apx{formdest}), so the level of fractionation in the three cyanides closely follows that of N/$^{15}$N (\fig{vcut-nratio}).

\begin{figure*}[t!]
\centering
\includegraphics[width=\hsize]{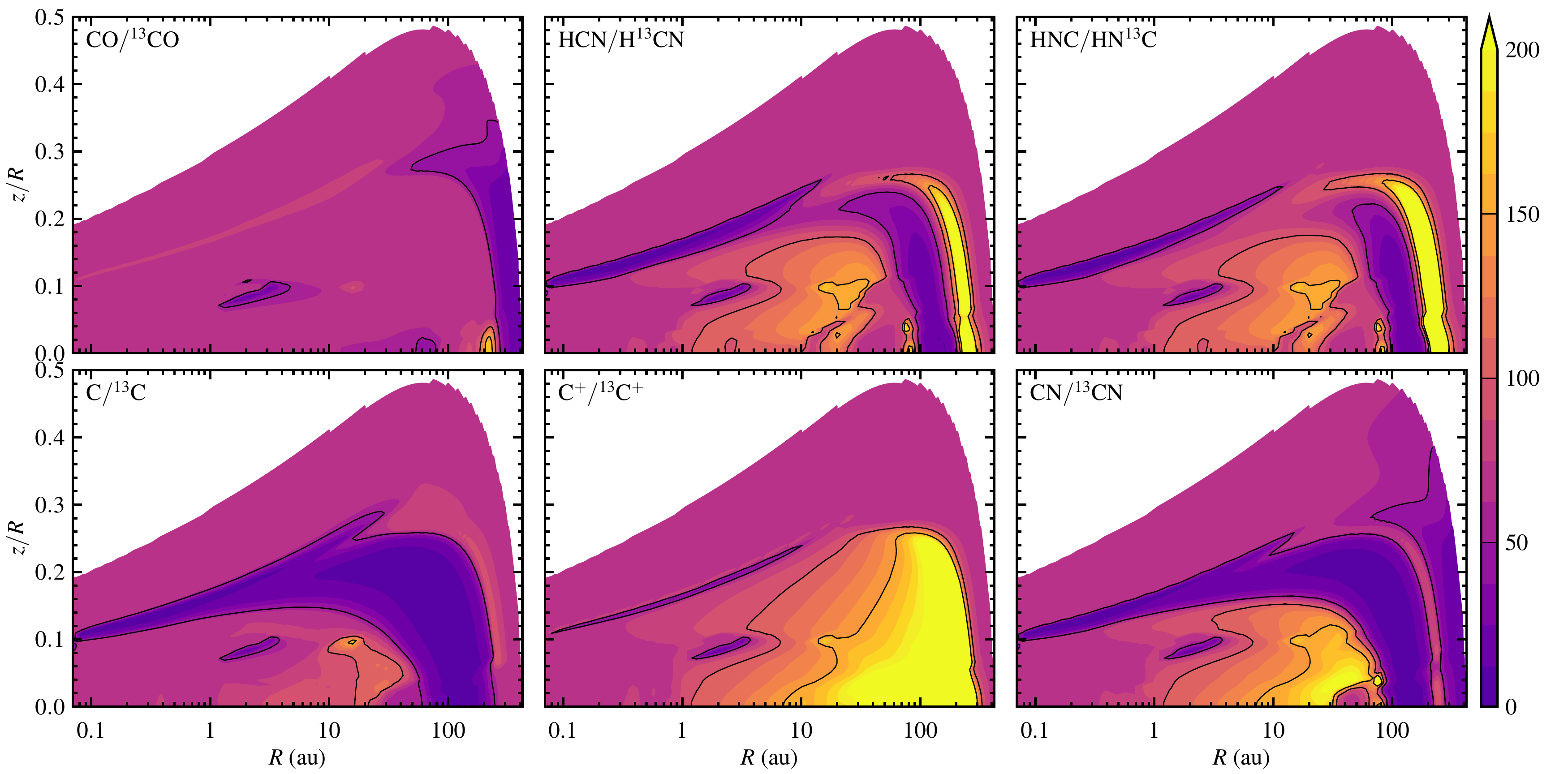}
\caption{Carbon isotope ratios of several molecules at 1 Myr for the same model as in \fig{physprop}. The elemental \ctw/\cth ratio is 69. Contours are drawn at ratios of 50, 100 and 150.}
\label{fig:ratio6c}
\end{figure*}

\begin{figure*}[t!]
\centering
\includegraphics[width=\hsize]{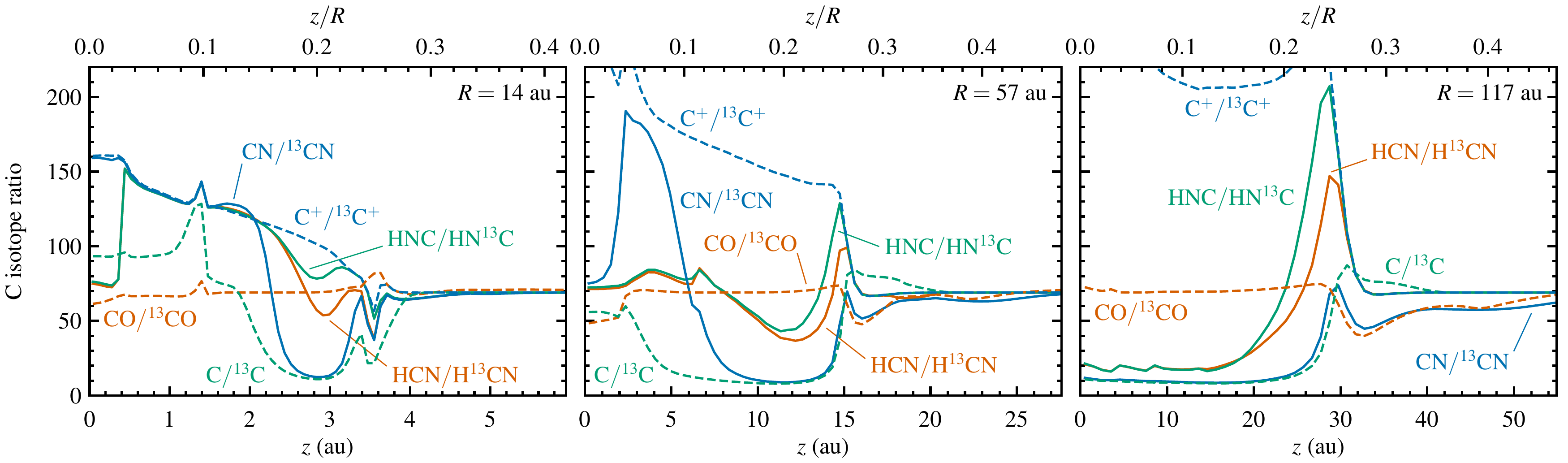}
\caption{Vertical cuts corresponding to \fig{ratio6c}.}
\label{fig:vcut-cratio}
\end{figure*}

Lastly, ammonia shows both \nc enhancement and depletion. The enhancement is strongest at the midplane around 170 au, with an \amh/$^{15}$NH$_3$ ratio down to 94. Depletion of \nc is seen in a layer from $z/R=0.10$ at the inner edge to $z/R=0.22$ at 40 au and curving down to $z/R=0.06$ near 150 au. The highest \amh/$^{15}$NH$_3$ ratio in the model is 800, corresponding to \nc depletion by a factor of 1.8.

The dual behavior of the \amh/$^{15}$NH$_3$ ratio is due to different sources of the nitrogen atom in ammonia. Throughout the disk, \amh is formed by successive hydrogenation of NH$^+$ and dissociative recombination of the resulting NH$_4^+$. The difference lies in the origin of NH$^+$. In the region of reduced \amh/$^{15}$NH$_3$ (i.e., \nc enhancement), the primary formation pathway of NH$^+$ is photoionization of NH\@. In turn, NH arises from atomic N and vibrationally excited \mh via N + H$_2^*$ $\to$ NH + H (\rx{n/h2*}). Hence, the isotope fractionation in \amh follows that of atomic N\@. In the other region, where ammonia is depleted in \nc, NH$^+$ is formed in two steps: \mn reacts with He$^+$ to N and N$^+$, and the latter reacts with \mh to NH$^+$. Since \amh is ultimately formed from \mn in these two regions, the \amh/$^{15}$NH$_3$ ratio follows the \mn/N$^{15}$N ratio.


\subsubsection{$^\mathsf{12}$C/$^\mathsf{13}$C}
\label{sec:12c13c}
The low-$J$ rotational lines of HCN are optically thick in most circumstellar disks. The \nqcr is therefore usually derived from \htcn and \hcnc, assuming a \ctw/\cth ratio of 70 \citep{hilyblant13a,wampfler14a,guzman15a,guzman17a}. Our model includes \cth in order to test if the assumption of a constant HCN/\htcn ratio is correct.

\figg{ratio6c} shows the carbon isotope ratios for CO, C, C$^+$, HCN, HNC and CN in the fiducial disk model, and the corresponding vertical cuts appear in \fig{vcut-cratio}. The nitrogen isotope patterns in \fig{ratio6n} are relatively simple, because isotope-selective photodissociation is the only active fractionation mechanism. For carbon, both selective photodissociation and low-temperature isotope exchange are important in the overall isotope balance. Hence, the fractionation patterns of carbon are more complex than those of nitrogen.

Carbon fractionation is largely controlled by CO, the most abundant carbon species. At the outer edge of the disk, the column density of CO is insufficient to provide self-shielding against the interstellar radiation field or the scattered stellar UV light. Because the gas is cold ($<$30 K), the isotope exchange reaction $^{13}{\rm C}^+ \rxplus {\rm CO}\ \rightleftarrows\ {\rm C}^+ \rxplus ^{13}{\rm CO} \rxplus 35\ {\rm K}$ is faster in the forward direction than backwards. This leads to an enhancement in \thco, visible as CO/\thco ratios down to 12 in the top left panel of \fig{ratio6c}.

Moving in from the outer edge, \twco becomes self-shielded once it builds up a column density of $\sim$\ten{14} \pcs. From here on in, \thco photodissociates faster than \twco\@. The gas temperature increases at the same time (from $\sim$20 to $\sim$30 K), reducing the effect of the low-temperature exchange process. The net result is an extended region with a minor increase in the CO/\thco ratio of up to 80 and a smaller spot near the midplane at 220 au with CO/\thco up to 170. The region of modest \thco depletion runs from $z/R=0.04$ at $R=220$ au up and in to $z/R=0.26$ at 100 au, and continues all along the disk surface to $z/R=0.1$ at the inner edge.

The fractionation pattern for atomic C is opposite to that of CO, not only spatially, but also in magnitude. The region of strong \thco enhancement at the outer edge corresponds to moderate depletion (up to 40\%) in atomic \cth, while the extended region of modest \thco depletion shows strong enhancement of \cth (up to a factor of 9). In all cases, photodissociation of CO and \thco is the major source of C and \cth\@. The outer edge of the disk is where C and C$^+$ take over from CO as the dominant carbon carriers, so even a large change in the CO/\thco ratio has little effect on the isotope ratio of the much more abundant atomic species. The reverse effect is seen away from the outer edge, where most carbon is locked up in CO: a small change in the photodissociation rates of CO and \thco gets amplified into a large difference in the C/\cth abundance ratio.

Ionized carbon shows no depletion at the very outer edge of disk, where its high abundance leaves it unaffected by the fractionation of CO\@. The bulk of the disk has an elevated C$^+$/$^{13}$C$^+$ ratio (up to 920) and that ratio generally increases with decreasing gas temperature. The responsible mechanism is the aforementioned reaction of $^{13}$C$^+$ with CO\@. Even though $^{13}$C$^+$ is formed from a reservoir of enhanced $^{13}$C, the exchange with CO has a larger effect and causes strong depletion of $^{13}$C$^+$. However, there is one region with $^{13}$C$^+$ enhancement: a thin layer along the disk surface, from $R=0.1$ to 10 au. This is at the CO/C/C$^+$ transition, where the gas temperatures of a few 100 K are too high for the low-temperature exchange mechanism to still be active. Instead, the $^{13}$C/$^{13}$C$^+$ ratio is set by the isotope-selective photodissociation of CO.

At the outer edge of the disk ($R>200$ au), CN is closely linked to and follows the fractionation of CO\@. Photodissociation and ionization convert CO to C$^+$, which rapidly reacts with NH\@. The product CN$^+$ then undergoes charge exchange with H to form CN\@. The CO--CN loop is closed by the reaction between CN and O to form CO and N\@. HCN and HNC are about four orders of magnitude less abundant than CN and are not closely linked to it. Instead, HCN and HNC are mostly formed by the dissociative recombination of HCNH$^+$, which in turn arises from C$^+$ reacting with \amh. The ratios of HCN/\htcn and HNC/H$^{13}$NC in this part of the disk follow C$^+$/$^{13}$C$^+$, i.e., no fractionation at the very outer edge and \cth depletion around $R=200$--300 au.

Moving in to around 100 au, all three cyanide species show \cth enhancements of up to a factor of 6. Due to the lower abundance of C$^+$, formation of CN is now dominated by the reaction of C with NH or NO\@. Neutral-neutral chemistry also controls the abundance HCN, in particular through the reaction of CH with NO\@. Charge exchange reactions through the intermediate HCNH$^+$ allow both CN and HCN to be converted to HNC\@. CH is formed from C and \mhstar, so the carbon fractionation in all three cyanides follows the fractionation of atomic C\@. The same coupling between C and the cyanides is also seen in the surface layers.

\begin{figure}[t!]
\centering
\includegraphics[width=\hsize]{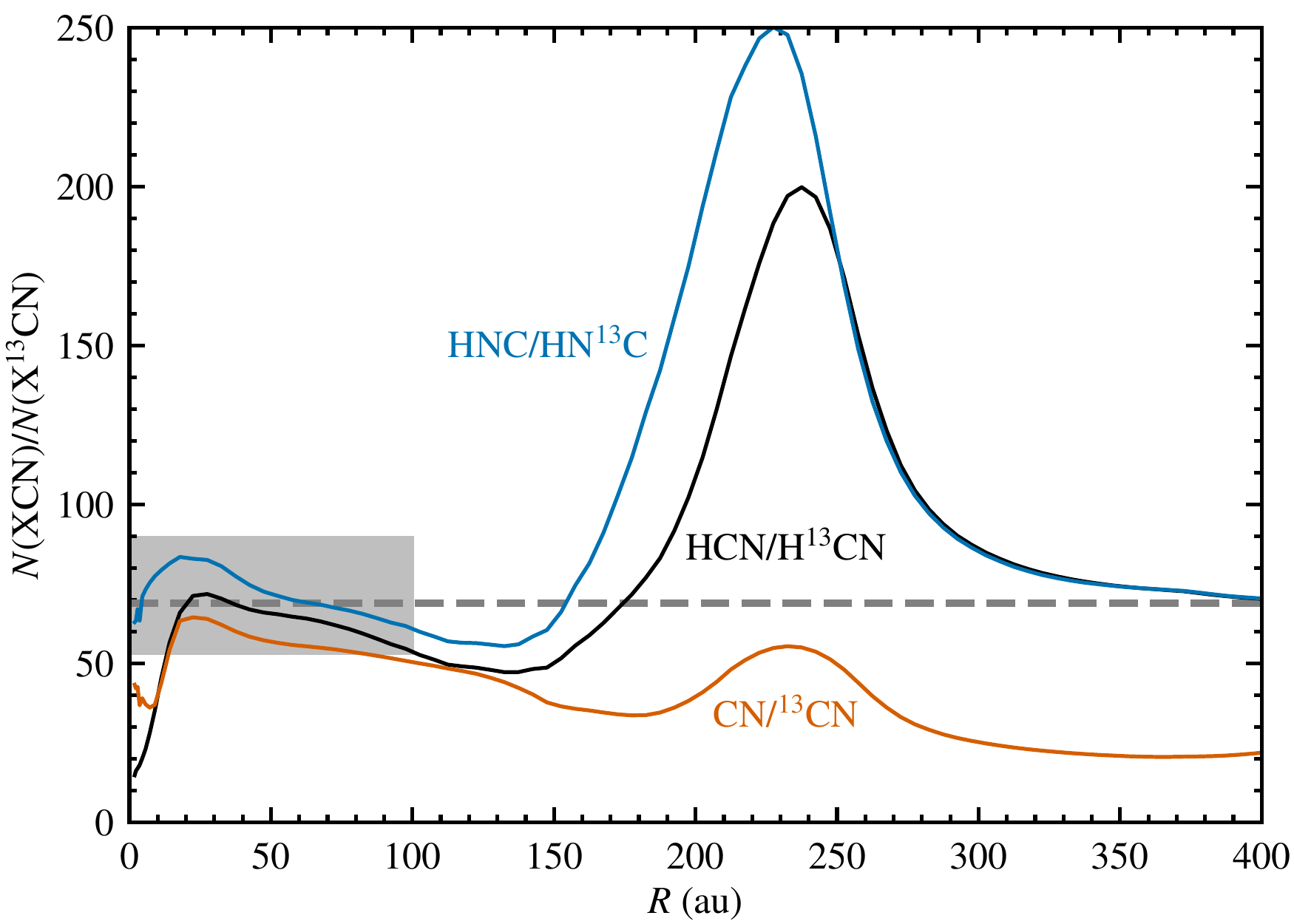}
\caption{Column density ratios of \ctw/\cth in HCN, HNC and CN at 1 Myr as a function of radius in the fiducial disk model. The dashed gray line marks the elemental \ctw/\cth ratio of 69. The shaded region indicates the 30\% uncertainty assumed by \citet{guzman17a}, whose observations probed material out to 100 au.}
\label{fig:colratio-hcn-h13cn}
\end{figure}

In order to compare with observations, column density ratios are more useful than abundance ratios. \figg{colratio-hcn-h13cn} shows the ratio of $N({\rm HCN})$ over $N({\rm H}^{13}{\rm CN})$ as a function of radius, ranging from a global minimum of 15 at 1 au and a local minimum of 47 at 130 au to a maximum of 200 at 240 au. The observations of \citet{guzman17a} probed \htcn and \hcnc out to radii of 100 au with a beam size of about 60 au. Under these conditions, \fig{colratio-hcn-h13cn} essentially shows a constant \ctw/\cth ratio in HCN\@. The variations in the inner 100 au are smaller than the uncertainty of 30\% assumed by \citeauthor{guzman17a} to convert \htcn column densities to HCN column densities. If future observations probe gas out to larger radii, the assumption of a constant \ctw/\cth ratio will need to be reevaluated.

\section{Discussion}
\label{sec:disc}

\subsection{Dominant isotope fractionation mechanism and comparison with observations}
\label{sec:mech}
\begin{figure*}[t!]
\centering
\includegraphics[width=\hsize]{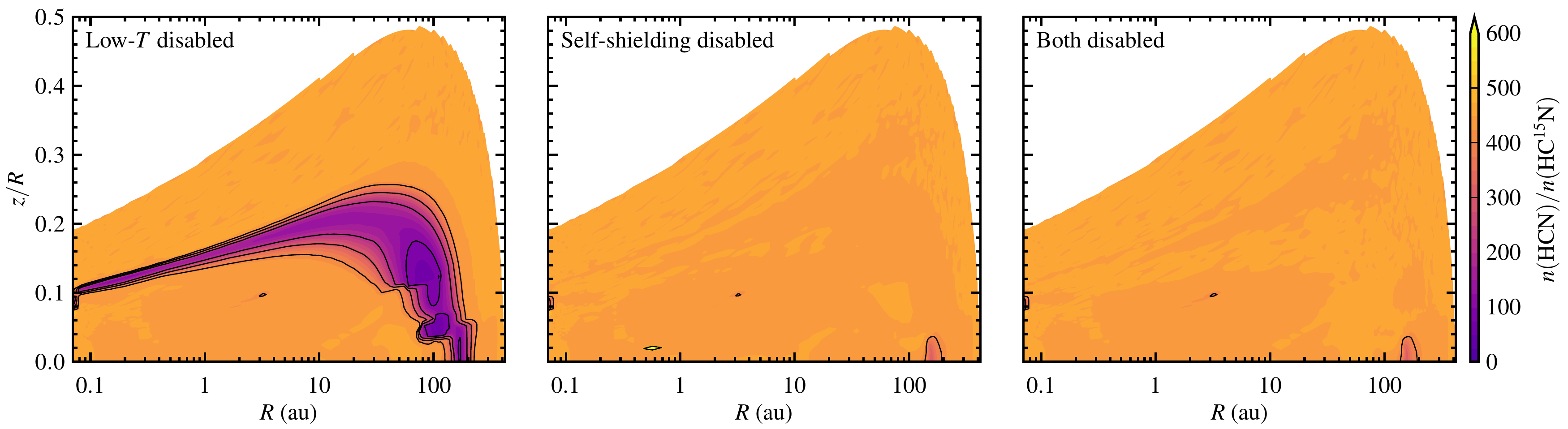}
\caption{Abundance ratios of HCN over \hcnc at 1 Myr in the fiducial disk model for three variations of the chemical network: with low-temperature isotope exchange reactions disabled (left), with self-shielding of \mn disabled (center), or with both fractionation mechanisms disabled (right). The ratios obtained with both mechanisms active are shown in the top center panel of \fig{ratio6n}. Contours are drawn at ratios of 100, 200, 300, 400 and 500.}
\label{fig:mechratio}
\end{figure*}

 We reran the fiducial model with either the low-temperature isotope exchange reactions turned off or with the self-shielding of \mn turned off, and once more with both mechanisms turned off. The left and center panels of \fig{mechratio} show the effects of enabling only one fractionation mechanisms at a time. Starting from the original ratios, with both mechanisms switched on (\fig{ratio6n}, top center), the ratios do not change when the low-temperature exchange reactions are disabled (\fig{mechratio}, left). The ratios do change substantially when self-shielding is turned off (\fig{mechratio}, middle), indicating that self-shielding of \mn is the dominant fractionation mechanism in our models.

\begin{figure}[t!]
\centering
\includegraphics[width=\hsize]{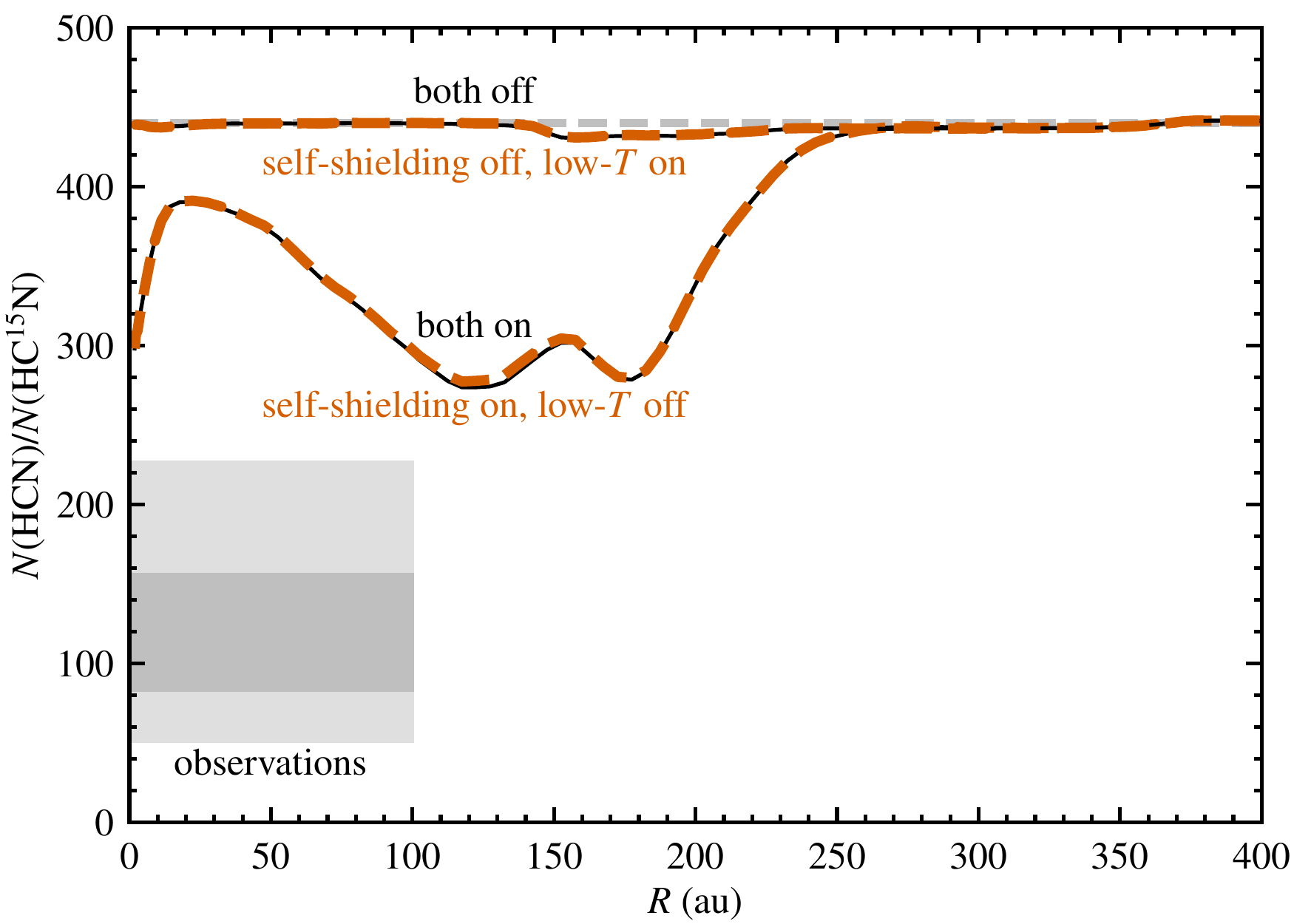}
\caption{Column density ratio of HCN/\hcnc at 1 Myr as a function of radius in the fiducial disk model for four variations of the chemical network: low-temperature exchange reactions on or off and self-shielding of \mn on or off. The dashed gray line marks the elemental \nqcr of 440. The dark and light gray shaded regions indicate the range of observed ratios and uncertainties from \citet{guzman17a}, probing material out to 100 au.}
\label{fig:mechcol}
\end{figure}

Another way to demonstrate the importance of self-shielding over low-$T$ fractionation is to plot column density ratios (\fig{mechcol}). The curves for ``both mechanisms off'' and ``self-shielding off, low-$T$ on'' fall on top of each other, as do the curves for ``both on'' and ``self-shielding on, low-$T$ off''. Hence, the column density ratios are completely unaffected by whether or not the low-temperature pathways are active.

As noted by \citet{wirstrom18a}, the reaction rates for some of the low-$T$ exchange reactions may have been underestimated by \citet{roueff15a}. Given that the bulk of the disk is warmer than 20 K, it appears that our conclusion regarding the dominant fractionation mechanism is robust even against order-of-magnitude variations in the exchange rates.

The HCN/\hcnc abundance and column density ratios at a chemical age of 1 Myr do not depend on the initial ratios. Our fiducial model starts with $n({\rm HCN})/n({\rm HC}^{15}{\rm N})=440$. We ran test models starting with ratios of 300 and 150, while keeping the elemental \nqcr at 440. In the absence of self-shielding, both models end up with the upper column density ratio profiles from \fig{mechcol}. With self-shielding turned on, the column density ratios match the lower curves. Hence, inheritance alone is not sufficient to explain the isotopolog ratios observed in disks. Instead, the observations require active fractionation powered by self-shielding of \mn.

With self-shielding active, the fiducial model produces abundance ratios from 55 to 440 for both HCN/\hcnc and CN/\cnc\@. Fractionation is strongest at radii from 50 to 200 au and altitudes from 40 au down to the midplane. No nitrogen isotope fractionation is seen at the midplane inside of 100 au, where comets and other solid bodies would be formed. Nonetheless, solar system comets show strong and uniform fractionation in both HCN and CN \citep[average \nqcrs of 140--150;][]{jehin09a,bockelee15a,hilyblant17a}. A combination of radial and vertical mixing can bring highly fractionated material into the comet-forming zone. Such mixing is also part of the theory of how isotope-selective photodissociation of CO resulted in the oxygen isotope ratios observed in meteorites \citep{lyons05a}.

The HCN/\hcnc column density ratio in the fiducial model varies between 270 and 440 in the inner 250 au. The lowest values appear between 100 and 200 au. The \htcn and \hcnc observations of six disks by \citet{guzman17a} probed gas at radii up to 100 au. The inferred HCN/\hcnc ratios range from $83\pm32$ to $156\pm71$ and are shown in \fig{mechcol} as shaded gray regions. Ignoring the uncertainties, the average ratio from the observed sample is a factor of 2 lower than the lowest column densities in the model and a factor of 3 lower than the average model ratio inside 100 au. Including uncertainties, the models and observations agree to better than a factor of 2.

With model HCN/\hcnc abundance ratios down to 55, the discrepancy in column density ratios is not due to insufficient levels of fractionation as such. Rather, the regions with strong fractionation do not coincide with the regions that contribute most to the column densities of HCN and \hcnc. Various trial runs have shown that the HCN morphology is sensitive to the treatment of vibrationally excited \mh. The current single-level approximation (\apx{h2star}) may not be appropriate for the cyanide chemistry, and a more detailed multi-level approach could be considered in the future. Furthermore, a proper comparison between observations and models requires full radiative transfer and excitation to compare actual line intensities. \citet{guzman17a} assumed local thermodynamic equilibrium, a single excitation temperature of 15 K and optically thin emission. These assumptions remain to be validated in future studies.

Throughout the disk, the abundance ratios of CN/\cnc are very similar to those of HCN/\hcnc (\fig{vcut-nratio}). In terms of column densities, however, CN shows less isotope fractionation than HCN. The $N({\rm CN})/N({\rm C}^{15}{\rm N})$ ratio in the fiducial model reaches a minimum of 310 at a radius of 180 au and shows an average of 400 in the inner 70 au. \citet{hilyblant17a} inferred a CN/\cnc ratio of $323\pm30$ on the same spatial scales in the disk around the T Tauri star TW Hya. Future observations will hopefully tell whether this ratio is unique to this source or representative of circumstellar disks in general.\footnote{\textit{Note added in proof:} Our adopted elemental \nqcr of 440 was also used in earlier models \citep[e.g.,][]{wirstrom12a,roueff15a}. Some observations suggest a ratio of about 300 for the present-day solar neighborhood \citep[e.g.][]{ritchey15a,hilyblant17a}. If our models were to use such an elemental ratio, the abundance and column density ratios of HCN/\hcnc, CN/\cnc, etc. would be lower by about a third. Isotope-selective photodissociation would remain the dominant fractionation mechanism.}


\subsection{Effect of disk parameters}
\label{sec:param}
The results presented so far were for one particular disk model, with a gas mass of \ten{-3} $M_\odot$, 90\% large grains and a scale height angle exponent of $\psi=0.2$ (\sect{physmod}). The stellar radiation field was that of a 4000 K blackbody (representing a T Tauri star) with a UV excess for an accretion rate of \ten{-8} $M_\odot$ \per{yr}. The current subsection discusses how changing these parameters affects the fractionation of the nitrogen isotopes. All models herein include self-shielding of \mn and low-temperature isotope exchange reactions.

\begin{figure*}[t!]
\centering
\includegraphics[width=\hsize]{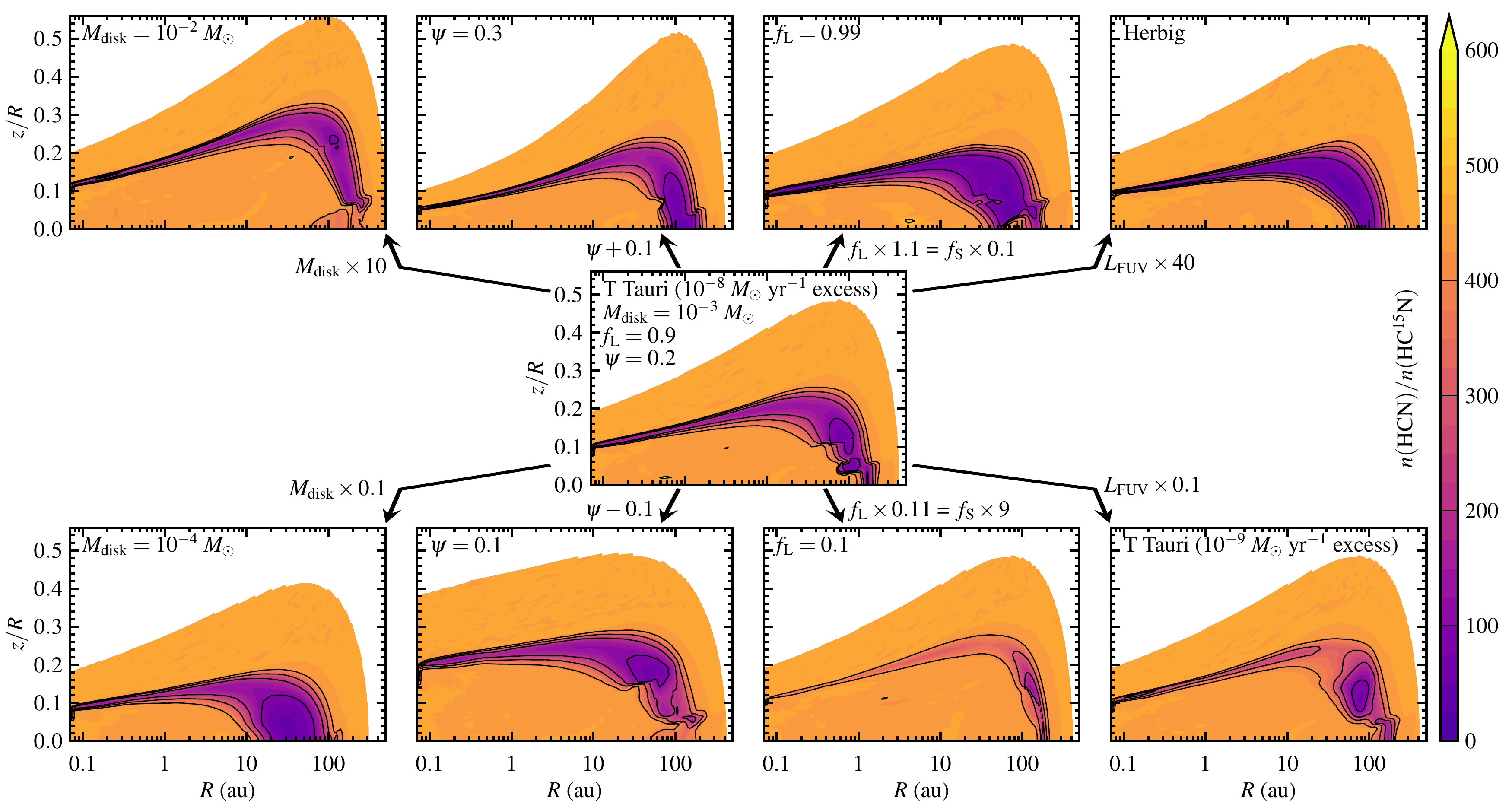}
\caption{Abundance ratios of HCN over \hcnc at 1 Myr for several sets of model parameters. The central panel is for the fiducial model from \sect{res} and is the same as the top center panel in \fig{ratio6n}. The other panels show the result of changing one parameter at a time, as indicated with the labels on the arrows and within the panels. Contours are drawn at ratios of 100, 200, 300, 400 and 500.}
\label{fig:pgabunratio}
\end{figure*}

\begin{figure*}[t!]
\centering
\includegraphics[width=\hsize]{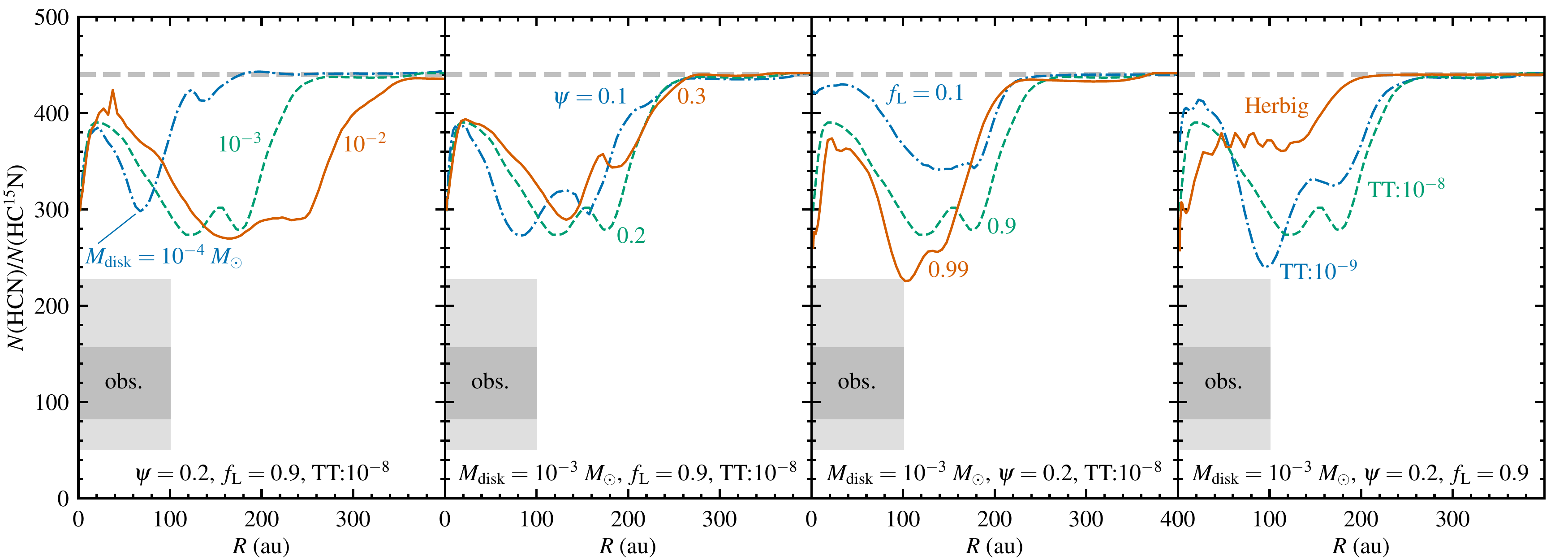}
\caption{Column density ratios of HCN over \hcnc at 1 Myr as a function of radius for a range of model parameters: disk mass ($M_{\rm disk}$), flaring angle power law ($\psi$), large grain fraction ($f_{\rm L}$) and stellar spectrum. The dashed gray line marks the elemental \nqcr of 440. The dark and light gray shaded regions indicate the range of observed HCN/\hcnc ratios and uncertainties from \citet{guzman17a}, probing material out to 100 au. The abbreviation ``TT:\ten{-8}'' means a T Tauri stellar spectrum with a UV excess for an accretion rate of \ten{-8} $M_\odot$ \per{yr}.}
\label{fig:pgridcolratio}
\end{figure*}

\figg{pgabunratio} shows how the HCN/\hcnc abundance ratios change when modifying one parameter at a time: disk mass, flaring angle, grain size distribution or stellar spectrum. \figg{pgridcolratio} shows the corresponding column density ratios. Despite obvious differences in both figures, the key point here is how little the column density ratios in the inner 100 au change across all model variations. The largest effect is seen when changing the grain size distribution, but even then the model ratios change by less than 50\%. The lack of variation in the model HCN/\hcnc column density ratios echoes the narrow range of values observed by \citet{guzman17a} in a sample of six disks with different characteristics, although the observed ratios are about a factor of 2 lower than the model ratios.

Increasing the disk mass (while keeping all other parameters constant) leads to higher densities at every point in the disk, and thus also to larger optical depths. Both formation and fractionation of HCN are governed by the UV field; the former through vibrational excitation of \mh, the latter through self-shielding of \mn. As excitation of \mh moves up to higher altitudes, so does the layer with high cyanide abundances. Likewise, as \mn becomes self-shielded at greater $z/R$, the region of strong isotope fractionation also moves up. The net result is no significant change in the column density ratios of HCN/\hcnc in the inner 100 au (\fig{pgridcolratio}, first panel).

Increasing the scale height angle exponent leads to a more strongly flared disk, where the outer parts intercept more stellar radiation than in a flatter disk. The layer with high HCN abundances and the layer with strong fractionation both move down in the inner disk and remain at the same altitude in the outer disk. The effects on $N({\rm HCN})/N({\rm HC}^{15}{\rm N})$ in the inner 100 au are again small (\fig{pgridcolratio}, second panel).

Bigger changes in the column density ratios are seen when changing the grain size distribution, as already noted by \citet{heays14a}. Decreasing the fraction of large grains from 90\% to 10\% leads to an increase in grain surface area per unit gas density. Hence, for the same column density of gas, the UV flux is attenuated more strongly by the dust. This reduces the importance of self-shielding, resulting in lower levels of isotope fractionation (\fig{pgabunratio}, bottom row, third panel). The model with $f_{\rm L}=0.1$ therefore has a higher HCN/\hcnc column density ratio than the fiducial model with $f_{\rm L}=0.9$ (\fig{pgridcolratio}, third panel). Likewise, an increase in $f_{\rm L}$ from 0.9 to 0.99 results in somewhat smaller column density ratios.

The last parameter we varied is the stellar spectrum, as detailed in \sect{physmod}. The fiducial spectrum represents a T Tauri star with a UV excess for an accretion rate of \ten{-8} $M_\odot$ \per{yr} ($L_{\rm FUV}=0.018$ $L_\odot$ from 6 to 13.6 eV). Reducing the accretion rate by one order of magnitude has little effect on the HCN/\hcnc ratio (\fig{pgridcolratio}, last panel). The ratios also do not change significantly when going from the fiducial model to a stellar spectrum typical of a Herbig star ($L_{\rm FUV}=0.77$ $L_\odot$), consistent with the similar ratios observed in T Tauri and Herbig stars by \citet{guzman17a}. The external radiation field has an equally small effect on the column density ratios, even when increased to 100 times the normal value for the solar neighborhood.


\section{Conclusions}
\label{sec:conc}
This paper presents chemical models of nitrogen isotope fractionation in circumstellar disks with a 2D axisymmetric geometry. These are the first models to include both low-temperature isotope exchange reactions and isotope-selective photodissociation for nitrogen. Using the thermochemical code DALI, we derive abundance profiles and isotope ratios for several key nitrogen-bearing species and we study how these ratios depend on various disk parameters.

The formation of CN and HCN is powered by the endothermic reaction between vibrationally excited \mh and atomic N\@. \mh is excited primarily through UV pumping in the irradiated surface layers and outer parts of the disk. Hence, these are the regions where the cyanides are most abundant.

Nitrogen isotope fractionation is fully dominated by isotope-selective photodissociation of \mn. The low-temperature exchange reactions do not contribute at all. The lowest HCN/\hcnc abundance ratio in our fiducial model is 55, representing \nc enhancement by a factor of 8 relative to the elemental \nqcr of 440. The HCN/\hcnc column density ratio shows a minimum of 270 and a mean of 340 in the inner 100 au. Taking uncertainties into account, these ratios are consistent with ALMA observations to within a factor of two \citep{guzman17a}.

Because of optical depth issues, the \nqcr in HCN is usually derived from observations of \htcn and \hcnc. By also including \cth in our models, we tested to what extent the standard assumption of a constant HCN/\htcn ratio is justified. In the inner 100 au, probed by recent observations, the column density ratio of HCN to \htcn varies by less than 20\% when averaged over typical beam sizes. The assumption of a constant HCN/\htcn ratio is currently justified, but needs to be reevaluated if future observations probe gas at larger radii.

The observable level of fractionation in HCN is largely insensitive to variations in the disk mass, the flaring angle or the stellar radiation field, but it does depend on the grain size distribution. Larger grains have a smaller surface area per unit gas volume than do smaller grains. Grain growth therefore allows for more self-shielding of \mn relative to extinction by dust and thus produces stronger levels of nitrogen isotope fractionation.

An important remaining challenge is to explain the discrepancy between the HCN/\hcnc column density ratios in the models and those inferred from observations. Given the crucial role of vibrationally excited \mh in the chemistry of HCN, HNC and CN, the current single-level approximation (\apx{h2star}) may not be good enough. Future work could explore a more accurate multi-level treatment.


\begin{acknowledgements}
We thank the entire DALI team for fruitful discussions. Astrochemistry in Leiden is supported by the European Union A-ERC grant 291141 CHEMPLAN, by the Netherlands Research School for Astronomy (NOVA) and by a Royal Netherlands Academy of Arts and Sciences (KNAW) professor prize. All figures in this paper were made with the Python package \texttt{matplotlib} \citep{hunter07a}.
\end{acknowledgements}



\begin{appendix}

\section{Vibrationally excited H\textsubscript{2}}
\label{apx:h2star}


\subsection{Introduction}
In the disk's surface layers, the stellar UV field can pump molecular hydrogen into a range of vibrationally excited states. The excess vibrational energy is available to overcome reaction barriers \citep{stecher72a,tielens85a,agundez10a,bruderer12a}. This process is particularly important in our models for the endothermic reaction between N and \mh (\rx{n/h2*}). The product NH reacts with C or C$^+$ in one of the key routes to forming the C--N bond.

Following \citet{tielens85a}, DALI treats vibrationally excited \mh as a single chemical species. This \mhstar has a weighted average energy of 30\,163 K, corresponding to a vibrational pseudo-level of $v^\ast=6$ \citep{london78a}. The abundance of \mhstar is set by UV pumping, spontaneous decay and collisions with H and \mh. The UV pumping rate is 8 times the \mh photodissociation rate \citep{sternberg14a} and the spontaneous decay rate is \scit{2}{-7} \ps \citep{london78a}. Our current model differs from earlier DALI publications in the adopted collision rates.

DALI originally used the collision rates from Eq.\ A14 of \citet{tielens85a}, which represented fits to a small set of computational and experimental data from the 1960s and 70s. Given the importance of \mhstar for the cyanide abundances, we implemented new collision rates based on more recent and more extensive computational studies. Approximations remain necessary, as described below, because specific rates into or out of the $v=6$ level are not available in the literature.


\subsection{Collisions between H\textsubscript{2} and H}
\citet{lique15a} computed state-to-state rovibrational rates for collisions between \mh and atomic H\@. The rates are tabulated for gas temperatures from 100 to 5000 K and for \mh states from $v$,$J=0$,0 to 3,8. Ortho-para conversion is possible through reactive collisions. The 3,8 level has an energy of 22\,000 K above the ground state, so extrapolation to the $v^\ast=6$ pseudo-level at 30\,000 K is feasible.

For use in DALI, we need the overall collisional cooling or deexcitation rate from $v=6$ to 0, averaged over an unknown population of initial rotational states. In order to extrapolate from \citeauthor{lique15a}'s rates, first we compute the overall cooling rates from every upper level to the 0,0 or 0,1 ground state. The fastest way for an excited \mh molecule to cool down to the ground state is usually through a cascade of multiple $\Delta v$,$\Delta J$ transitions.

For example, the $v$,$J=0$,4 level can either undergo a single $\Delta J=-4$ collision or two $\Delta J=-2$ collisions. At 300 K, the rate coefficients from \citet{lique15a} are $k_{0402}=\scim{6.0}{-14}$ cm$^3$ \ps for 0,$4\to0$,2 and $k_{0200}=\scim{9.1}{-14}$ cm$^3$ \ps for 0,$2\to0$,0. The rate coefficient for the two $\Delta J=-2$ collisions together then becomes $(k_{0402}^{-1}+k_{0200}^{-1})^{-1}=\scim{3.6}{-14}$ cm$^3$ \ps. The single $\Delta J=-4$ collision has a rate coefficient of only \scit{2.9}{-15} cm$^3$ \ps, so the dominant deexcitation mechanism for the 0,4 level is through a mini-cascade of two $\Delta J=-2$ collisions.

\begin{figure}[t!]
\centering
\includegraphics[width=\hsize]{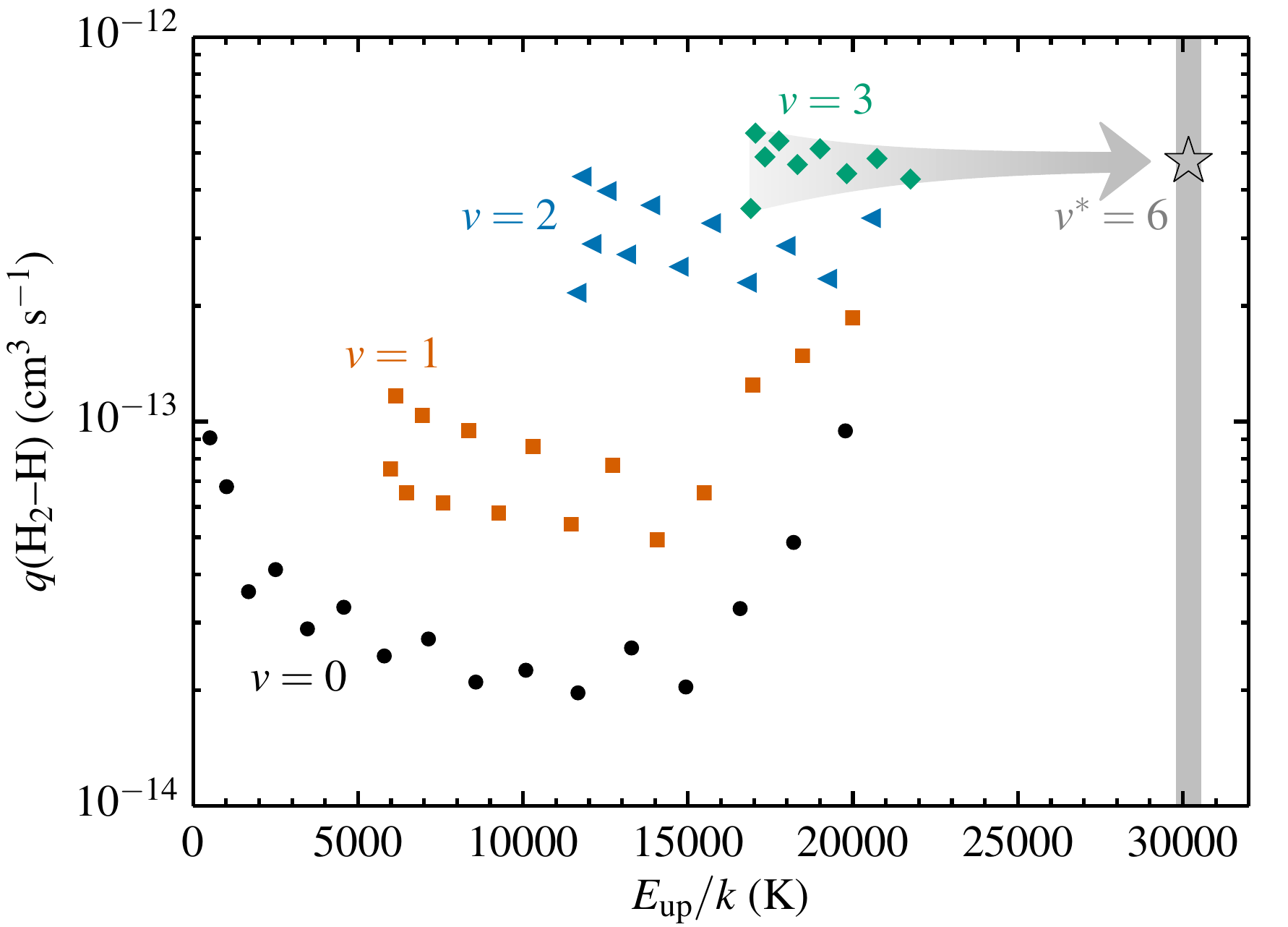}
\caption{Cooling or deexcitation rate coefficients for \mh colliding with H at a gas temperature of 300 K\@. Plotted are the overall rates from a given $v$,$J$ upper level to the 0,0 or 0,1 ground state, based on the fastest possible rovibrational cascade (see text). The rates are grouped by vibrational level: $v$,$J=0$,0 to 0,17 (black circles), 1,0 to 1,14 (orange squares), 2,0 to 2,11 (blue triangles) and 3,0 to 3,8 (green diamonds). The original state-to-state rates are from \citet{lique15a}. The gray star marks the adopted rate for our $v^\ast=6$ pseudo-level; as indicated by the arrow, this is the average from the $v=3$ rates.}
\label{fig:h2-h-coll}
\end{figure}

We repeat this analysis to find the fastest deexcitation cascade for each $v$,$J$ level in \citeauthor{lique15a}'s dataset. The corresponding rate coefficients at 300 K are plotted in \fig{h2-h-coll} as a function of upper-level energy. The rates are grouped by vibrational level: $v$,$J=0$,0 to 0,17 (black circles), 1,0 to 1,14 (orange squares), 2,0 to 2,11 (blue triangles) and 3,0 to 3,8 (green diamonds).

\begin{figure}[t!]
\centering
\includegraphics[width=\hsize]{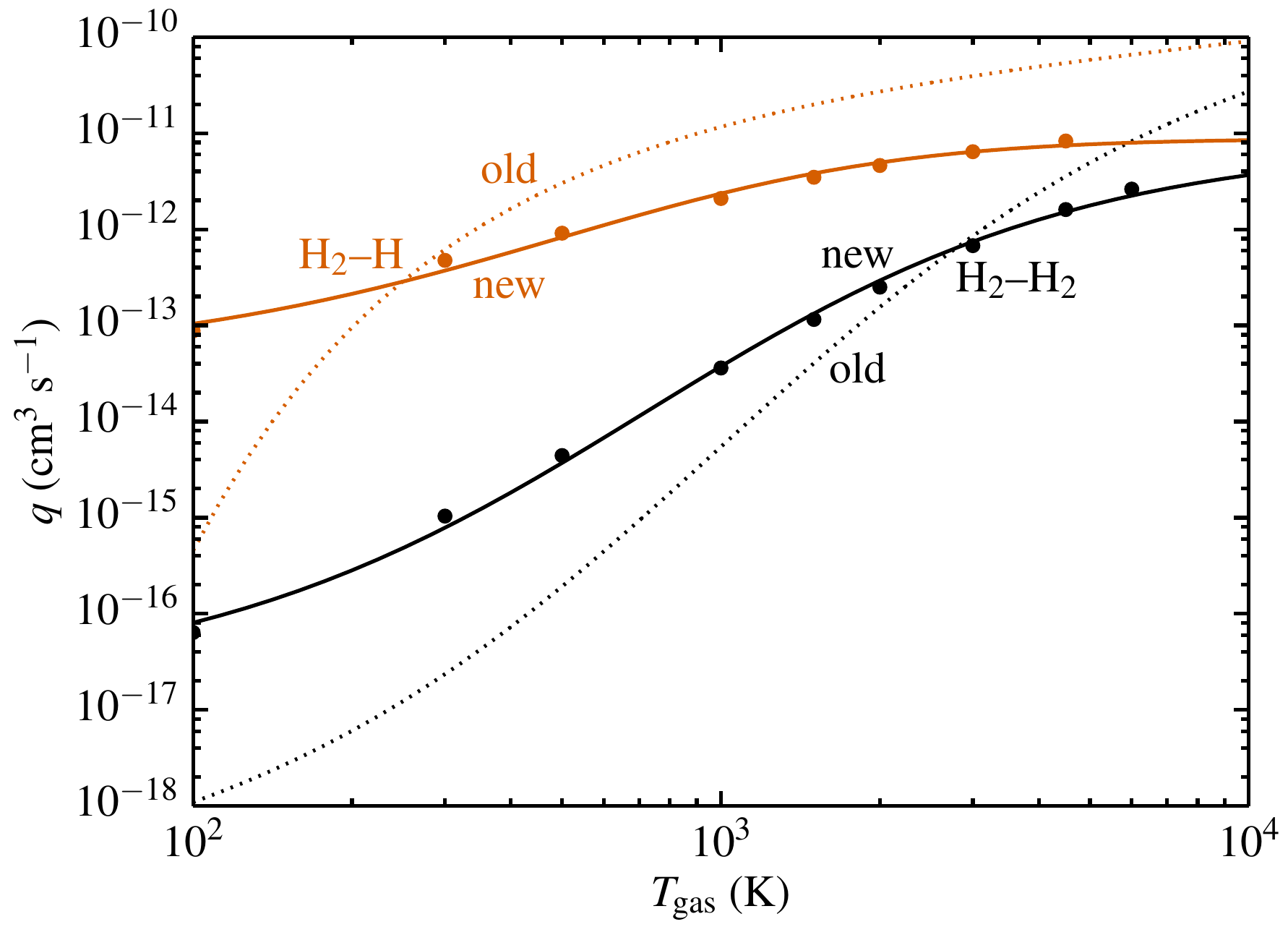}
\caption{Cooling or deexcitation rate coefficients as a function of gas temperature for \mh colliding with H (orange) or \mh (black). The ``old'' rates (dotted lines) are the original implementation in DALI, based on \citet{tielens85a}. The ``new'' rates (symbols and solid lines) are as described in this Appendix.}
\label{fig:collrates}
\end{figure}

The cooling rates increase from $v=0$ to 3. However, the difference between the $v=2$ and 3 rates is only 60\% at 300 K and disappears altogether at higher temperatures. The $v=3$ rates therefore appear to be a good proxy for the $v=6$ rates. Furthermore, there is little dependence on $J$ within the $v=3$ sets. At a given temperature, we thus take the average of the $v=3$ rates as the \mh--H collisional deexcitation rate for the $v=6$ pseudo-level. These rates are plotted versus temperature in \fig{collrates} and the 300 K rate is marked with a star in \fig{h2-h-coll}. For use in DALI, we fit the temperature dependence of the rates to the formula from \citet{lebourlot99a}:
\begin{equation}
\label{eq:qtfit}
\log \frac{q}{{\rm cm}^3\,{\rm s}^{-1}} = a + \frac{b}{t} + \frac{c}{t^2}\,,
\end{equation}
where $t=1+T/(1000\,{\rm K})$. The fitted coefficients for the \mh--H collisions are $a=-11.06$, $b=0.0555$ and $c=-2.390$.

\subsection{Collisions between H\textsubscript{2} and H\textsubscript{2}}
The other set of rates is for collisions between \mh and \mh. \citet{flower98a} computed state-to-state rovibrational rates up to $v$,$J=3$,8 at gas temperatures from 100 to 6000 K, covering nearly the same parameter space as \citet{lique15a} for \mh--H\@. \citet{fonseca13a} computed rates only up to the 2,4 level ($\eup/k=13\,000$ K) and for a narrower range of temperatures. \citet{ahn07a} performed experiments up to $v=5$ at a gas temperature of 300 K and used scaled semi-classical computations to estimate rates up to $v=9$. \citeauthor{flower98a} kept the collision partner fixed in the ground state, whereas \citeauthor{fonseca13a} and \citeauthor{ahn07a} allowed both \mh molecules to undergo rovibrational transitions. These so-called vibration-vibration (VV) collisions are three to four orders of magnitude faster than the vibration-translation (VT) collisions from \citeauthor{flower98a}.

\begin{figure}[t!]
\centering
\includegraphics[width=\hsize]{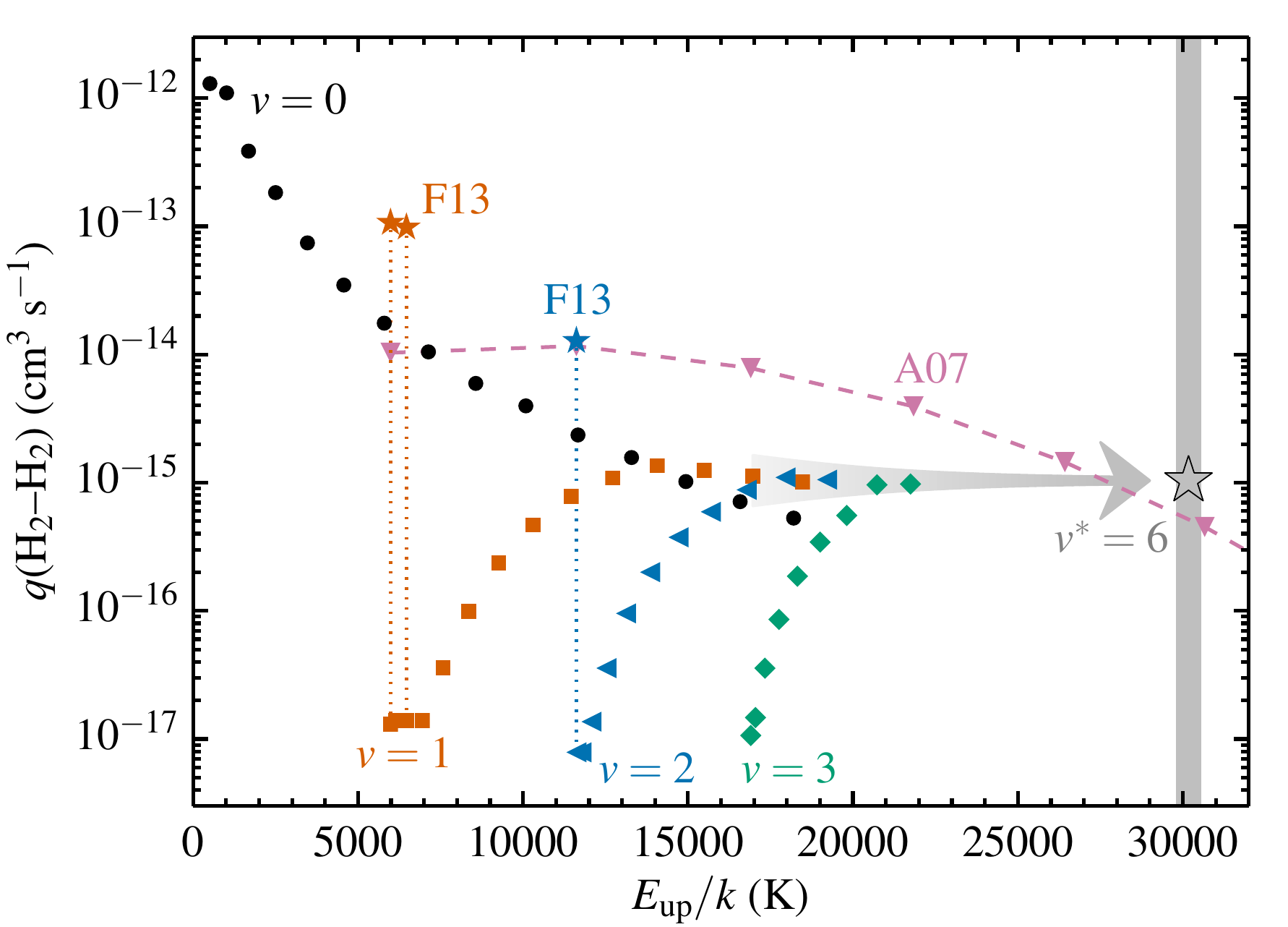}
\caption{Same as \fig{h2-h-coll}, but for \mh colliding with \mh. The circles, squares, triangles and diamonds denote the overall cooling rate coefficients from a given $v$,$J$ level to the ground state based on the state-to-state rates from \citet{flower98a}. The orange and blue stars show the $v$,$J=1$,0, 1,2 and 2,0 vibration-vibration rates of \citet{fonseca13a}. The lilac triangles and dashed curve show the VV rates of \citet{ahn07a}. The gray star marks the adopted rate for our $v^\ast=6$ pseudo-level; as indicated by the arrow, this is the rate that the $v=1$, 2 and 3 series from \citeauthor{flower98a} converge to in the high-$J$ limit.}
\label{fig:h2-h2-coll}
\end{figure}

As was the case for the \mh--H collisions, the overall deexcitation process from a given $v$,$J$ level typically involves multiple \mh--\mh collisions. We followed the same approach as above in finding the fastest cascade to compute overall cooling rates as a function of gas temperature and upper level. \figg{h2-h2-coll} shows the results for 300 K based on the state-to-state rates from \citet{flower98a}. The three VV rates tabulated by \citet{fonseca13a} are plotted as small stars and are connected to the corresponding VT rates from \citeauthor{flower98a} by dotted lines. The lilac triangles and dashed curve represent the rates from \citeauthor{ahn07a}

For the rate coefficients based on \citeauthor{flower98a}, the $v=1$, 2 and 3 series all converge to a common value of \scit{1}{-15} cm$^3$ \ps. A similar level of convergence occurs at all other temperatures in their grid. The $v=6$ estimate at 300 K from \citeauthor{ahn07a} is \scit{5}{-16} cm$^3$ \ps, a factor of 2 lower than the ``convergence rate'' derived from \citeauthor{flower98a}. The difference between the VV and VT rates thus appears to vanish at higher $v,J$ levels than those explored by \citeauthor{fonseca13a}

The study of \citeauthor{ahn07a} was limited to a single temperature of 300 K\@. For other temperatures, the best we can do is to adopt the rates that the $v=1$, 2 and 3 series from \citeauthor{flower98a} converge to. These rates are plotted in \fig{collrates}. In order to maintain a smooth temperature dependence, we also use the rate from \citeauthor{flower98a} at 300 K. The temperature dependence is fitted to \eq{qtfit} with $a=-11.08$, $b=-3.671$ and $c=-2.023$.


\section{List of species}
\label{apx:speclist}
\tb{speclist} lists the full set of species in our chemical network with carbon and nitrogen isotopes. The network contains 205 gas-phase species and 76 ices, connected by 5751 reactions.

\begin{table*}[t!]
\caption{List of species in our chemical network with carbon and nitrogen isotopes.}
\label{tb:speclist}
\centering
\begin{tabular}{ccccccccc}
\hline\hline
H & He & C & $^{13}$C & N & $^{15}$N & O & Mg & Si \\
S & Fe & H$_2$ & H$_2^*$ & CH & $^{13}$CH & CH$_2$ & $^{13}$CH$_2$ & CH$_3$ \\
$^{13}$CH$_3$ & CH$_4$ & $^{13}$CH$_4$ & NH & $^{15}$NH & NH$_2$ & $^{15}$NH$_2$ & NH$_3$ & $^{15}$NH$_3$ \\
OH & H$_2$O & C$_2$ & C$^{13}$C & C$_2$H & $^{13}$CCH & C$^{13}$CH & C$_2$H$_2$ & HC$^{13}$CH \\
C$_2$H$_3$ & H$^{13}$CCH$_2$ & HC$^{13}$CH$_2$ & CN & C$^{15}$N & $^{13}$CN & $^{13}$C$^{15}$N & HCN & HC$^{15}$N \\
H$^{13}$CN & H$^{13}$C$^{15}$N & HNC & H$^{15}$NC & HN$^{13}$C & H$^{15}$N$^{13}$C & H$_2$CN & H$_2$C$^{15}$N & H$_2^{13}$CN \\
H$_2^{13}$C$^{15}$N & CO & $^{13}$CO & HCO & H$^{13}$CO & H$_2$CO & H$_2^{13}$CO & CO$_2$ & $^{13}$CO$_2$ \\
N$_2$ & N$^{15}$N & NO & $^{15}$NO & HNO & H$^{15}$NO & O$_2$ & C$_2$N & C$_2^{15}$N \\
$^{13}$CCN & $^{13}$CC$^{15}$N & C$^{13}$CN & C$^{13}$C$^{15}$N & OCN & OC$^{15}$N & O$^{13}$CN & O$^{13}$C$^{15}$N & SiH \\
SiO & e$^-$ & H$^+$ & H$^-$ & H$_2^+$ & H$_3^+$ & He$^+$ & HeH$^+$ & C$^+$ \\
$^{13}$C$^+$ & CH$^+$ & $^{13}$CH$^+$ & CH$_2^+$ & $^{13}$CH$_2^+$ & CH$_3^+$ & $^{13}$CH$_3^+$ & CH$_4^+$ & $^{13}$CH$_4^+$ \\
CH$_5^+$ & $^{13}$CH$_5^+$ & N$^+$ & $^{15}$N$^+$ & NH$^+$ & $^{15}$NH$^+$ & NH$_2^+$ & $^{15}$NH$_2^+$ & NH$_3^+$ \\
$^{15}$NH$_3^+$ & NH$_4^+$ & $^{15}$NH$_4^+$ & O$^+$ & OH$^+$ & H$_2$O$^+$ & H$_3$O$^+$ & C$_2^+$ & C$^{13}$C$^+$ \\
C$_2$H$^+$ & $^{13}$CCH$^+$ & C$^{13}$CH$^+$ & C$_2$H$_2^+$ & HC$^{13}$CH$^+$ & C$_2$H$_3^+$ & H$^{13}$CCH$_2^+$ & HC$^{13}$CH$_2^+$ & CN$^+$ \\
C$^{15}$N$^+$ & $^{13}$CN$^+$ & $^{13}$C$^{15}$N$^+$ & HCN$^+$ & HC$^{15}$N$^+$ & H$^{13}$CN$^+$ & H$^{13}$C$^{15}$N$^+$ & HNC$^+$ & H$^{15}$NC$^+$ \\
HN$^{13}$C$^+$ & H$^{15}$N$^{13}$C$^+$ & HCNH$^+$ & HC$^{15}$NH$^+$ & H$^{13}$CNH$^+$ & H$^{13}$C$^{15}$NH$^+$ & H$_2$NC$^+$ & H$_2^{15}$NC$^+$ & H$_2$N$^{13}$C$^+$ \\
H$_2^{15}$N$^{13}$C$^+$ & CO$^+$ & $^{13}$CO$^+$ & HCO$^+$ & H$^{13}$CO$^+$ & HOC$^+$ & HO$^{13}$C$^+$ & H$_2$CO$^+$ & H$_2^{13}$CO$^+$ \\
H$_3$CO$^+$ & H$_3^{13}$CO$^+$ & CO$_2^+$ & $^{13}$CO$_2^+$ & HCO$_2^+$ & H$^{13}$CO$_2^+$ & N$_2^+$ & N$^{15}$N$^+$ & N$_2$H$^+$ \\
$^{15}$NNH$^+$ & N$^{15}$NH$^+$ & NO$^+$ & $^{15}$NO$^+$ & HNO$^+$ & H$^{15}$NO$^+$ & O$_2^+$ & CNC$^+$ & C$^{15}$NC$^+$ \\
CN$^{13}$C$^+$ & C$^{15}$N$^{13}$C$^+$ & C$_2$N$^+$ & C$_2^{15}$N$^+$ & $^{13}$CCN$^+$ & $^{13}$CC$^{15}$N$^+$ & C$^{13}$CN$^+$ & C$^{13}$C$^{15}$N$^+$ & C$_2$NH$^+$ \\
C$_2^{15}$NH$^+$ & $^{13}$CCNH$^+$ & $^{13}$CC$^{15}$NH$^+$ & C$^{13}$CNH$^+$ & C$^{13}$C$^{15}$NH$^+$ & OCN$^+$ & OC$^{15}$N$^+$ & O$^{13}$CN$^+$ & O$^{13}$C$^{15}$N$^+$ \\
HNCO$^+$ & H$^{15}$NCO$^+$ & HN$^{13}$CO$^+$ & H$^{15}$N$^{13}$CO$^+$ & Mg$^+$ & Si$^+$ & SiH$^+$ & SiH$_2^+$ & SiO$^+$ \\
SiOH$^+$ & S$^+$ & Fe$^+$ & PAH$^0$ & PAH$^+$ & PAH$^-$ & PAH:H &  &  \\
\hline
\end{tabular}
\tablefoot{\sectt{chemnet} describes the network in full. H$_2^*$ is vibrationally excited \mh. C and N without superscripts denote $^{12}$C and $^{14}$N\@. PAH$^0$, PAH$^+$ and PAH$^-$ are neutral and ionized PAHs, and PAH:H is a hydrogenated PAH\@. All neutral species except PAHs are present in the gas phase and as ice; PAHs, e$^-$ and all ions are present in the gas only.}
\end{table*}


\section{Formation, conversion and destruction of HCN, HNC and CN}
\label{apx:formdest}
The cyanide abundances throughout the disk are dominated by just a few formation and destruction reactions. Each of the initial nitrogen reservoirs (\amh ice, N and \mn) can be converted into HCN, HNC and CN\@. The first such pathway starts with evaporation of \amh ice followed by proton transfer:
\begin{align}
{\rm NH}_3 \rxplus {\rm H}_3{\rm O}^+ \rxtoa {\rm NH}_4^+ \rxplus {\rm H}_2{\rm O}\, \label{rx:amh/h3o+} \\
{\rm NH}_4^+ \rxplus {\rm C} \rxtoa {\rm HCNH}^+ \rxplus {\rm H}_2 \nonumber \\
                                 &\ {\rm H}_2{\rm NC}^+ \rxplus {\rm H}_2\,, \label{rx:nh4+/c}
\end{align}
Proton transfer to \amh also occurs from HCO$^+$, H$_3^+$ and other ions. The products HCNH$^+$ and H$_2$NC$^+$ undergo proton transfer or recombine with an electron to produce HCN, HNC and CN \citep{loison14a}.

The second cyanide formation pathway involves atomic N and vibrationally excited \mh \citep{tielens85a,bruderer12a}:
\begin{align}
{\rm N} \rxplus {\rm H}_2^* \rxtoa {\rm NH} \rxplus {\rm H}\, \label{rx:n/h2*} \\
{\rm NH} \rxplus {\rm C} \rxtoa {\rm CN} \rxplus {\rm H}\, \label{rx:nh/c} \\
{\rm NH} \rxplus {\rm C}^+ \rxtoa {\rm CN}^+ \rxplus {\rm H}\,. \label{rx:nh/c+}
\end{align}
\rx{n/h2*} has a barrier of 12\,650 K \citep{davidson90a}, so the rate with ground-state \mh{} is negligibly slow below gas temperatures of $\sim$1200 K\@. The intermediate CN$^+$ from \rx{nh/c+} is hydrogenated to HCNH$^+$, which recombines to HCN, HNC and CN\@. A variation of the above scheme involves NH reacting with O to NO, followed by NO reacting with C to CN or with CH to HCN.

Of course, there are several other sets of reactions to form cyanides out of atomic nitrogen, such as
\begin{equation}
{\rm N} \rxplus {\rm CH}_2 \rxto {\rm HCN} + {\rm H}\, \label{rx:n/ch2}
\end{equation}
and
\begin{align}
{\rm N} \rxplus {\rm C}_2{\rm H} \rxtoa {\rm C}_2{\rm N} + {\rm H}\, \label{rx:n/c2h} \\
{\rm C}_2{\rm N} \rxplus {\rm N} \rxtoa {\rm CN} \rxplus {\rm CN}\, \label{rx:c2n/n}
\end{align}
and
\begin{align}
{\rm N} \rxplus {\rm OH} \rxtoa {\rm NO} + {\rm H}\, \label{rx:n/oh} \\
{\rm C} \rxplus {\rm NO} \rxtoa {\rm CN} \rxplus {\rm O}\, \label{rx:c/no} \\
{\rm CH} \rxplus {\rm NO} \rxtoa {\rm HCN} \rxplus {\rm O}\,. \label{rx:ch/no}
\end{align}
None are as important as the \mhstar route, but they all play a supporting role at various points in the disk.

The third and final set of formation routes starts from \mn:
\begin{align}
&{\rm N}_2 \rxplus {\rm He}^+ \rxto {\rm N} \rxplus {\rm N}^+ \rxplus {\rm He}\, \label{rx:n2/he+} \\
&{\rm N}^+\ \xrightarrow{{\rm H}_2}\ {\rm NH}^+\ \xrightarrow{{\rm H}_2}\ {\rm NH}_2^+\ \xrightarrow{{\rm H}_2}\ {\rm NH}_3^+\ \xrightarrow{{\rm H}_2}\ {\rm NH}_4^+\,, \label{rx:n+/h2}
\end{align}
where the intermediate products NH$_4^+$ and N feed into Reactions \ref{rx:nh4+/c}, \ref{rx:n/h2*}, \ref{rx:n/ch2}, \ref{rx:n/c2h} and \ref{rx:n/oh} to produce HCN, HNC and CN.

Once formed, the three cyanide species can be converted into each other via several sets of reactions. In the surface layers of the disk, photodissociation reduces HCN and HNC to CN:
\begin{align}
{\rm HCN} \rxplus h\nu \rxtoa {\rm CN} \rxplus {\rm H}\, \label{rx:hcn/uv} \\
{\rm HNC} \rxplus h\nu \rxtoa {\rm CN} \rxplus {\rm H}\,. \label{rx:hnc/uv}
\end{align}
The second set of conversion reactions starts with proton transfer to HCN or HNC (from e.g. HCO$^+$, H$_3$O$^+$ or H$_3^+$), followed by
\begin{align}
{\rm HCNH}^+ \rxplus {\rm NH}_3 \rxtoa {\rm HCN} \rxplus {\rm NH}_4^+ \nonumber \\
                                    &\ {\rm HNC} \rxplus {\rm NH}_4^+ \label{rx:hcnh+/nh3}
\end{align}
or
\begin{align}
{\rm HCNH}^+ \rxplus {\rm e}^- \rxtoa {\rm HCN} \rxplus {\rm H} \nonumber \\
                                   &\ {\rm HNC} \rxplus {\rm H} \nonumber \\
                                   &\ {\rm CN} \rxplus {\rm H}_2\,. \label{rx:hcnh+/e-}
\end{align}
The two or three products have roughly equal branching ratios \citep{loison14a}. Lastly, there are three important neutral-neutral conversion reactions:
\begin{align}
{\rm CN} \rxplus {\rm H}_2 \rxtoa {\rm HCN} \rxplus {\rm H}\, \label{rx:cn/h2} \\
{\rm HNC} \rxplus {\rm C}  \rxtoa {\rm HCN} \rxplus {\rm C}\, \label{rx:hnc/c} \\
{\rm HNC} \rxplus {\rm H}  \rxtoa {\rm HCN} \rxplus {\rm H}\,. \label{rx:hnc/h}
\end{align}
\rx{cn/h2} has an activation barrier of 960 K \citep{baulch05a}, while the rate of \rx{hnc/c} is constant with temperature \citep{loison14a}. \rx{hnc/h} uses the two-component fit from \citet{leteuff00a} to the quantum chemical computations of \citet{talbi96a}: $k=10^{-15}$ cm$^3$ \ps below 100 K and $k=\scim{1.36}{-13} (T/300\, {\rm K})^{4.48}$ cm$^3$ \ps above 100 K.

Although plenty of reactions exist that can break the C--N bond in HCN and HNC, none of these are dominant anywhere in our disk models. Instead, loss of the C--N bond occurs exclusively from CN through the following reactions:
\begin{align}
{\rm CN} \rxplus h\nu \rxtoa {\rm C} \rxplus {\rm N}\, \label{rx:cn/uv} \\
{\rm CN} \rxplus {\rm N} \rxtoa {\rm N}_2 \rxplus {\rm C}\, \label{rx:cn/n} \\
{\rm CN} \rxplus {\rm O} \rxtoa {\rm CO} \rxplus {\rm N}\, \label{rx:cn/o} \\
{\rm CN} \rxplus {\rm O}_2 \rxtoa {\rm CO} \rxplus {\rm NO}\, \nonumber \\
                               &\ {\rm OCN} \rxplus {\rm O}\,. \label{rx:cn/o2}
\end{align}
Cleaving the C--N bond in HCN or HNC requires conversion to CN via Reactions \ref{rx:hcn/uv}, \ref{rx:hnc/uv} and \ref{rx:hcnh+/e-}.

\end{appendix}

\end{document}